\input harvmac
\input epsf
\noblackbox
\Title{\vbox{
\hbox{HUTP-01/A011, UCSD-PTH-01-03}
\hbox{\tt hep-th/0103067}}}{A Large $N$ Duality via a Geometric
Transition}
\medskip
\centerline{F. Cachazo$^1$, K. Intriligator$^2$  and
C. Vafa$^1$}
\medskip
\centerline{$^1$ Jefferson Physical Laboratory}
\centerline{Harvard University}
\centerline{Cambridge, MA 02138, USA}
\medskip
\centerline{$^2$ UCSD Physics Department}
\centerline{9500 Gilman Drive}
\centerline{La Jolla, CA 92093}
\bigskip

\vskip .3in
We propose a large $N$ dual of 4d, ${\cal N}=1$ supersymmetric,
$SU(N)$ Yang-Mills with adjoint field $\Phi$ and arbitrary
superpotential $W(\Phi)$.  The field theory is geometrically
engineered via D-branes
partially wrapped over certain cycles of a non-trivial Calabi-Yau
geometry.  The large $N$, or low-energy, dual arises from a geometric
transition of the Calabi-Yau, where the branes have disappeared and
have been replaced by suitable fluxes.  This duality yields highly
non-trivial exact results for the gauge theory.  The predictions indeed
agree with expected results in cases where it is possible to use
standard techniques for analyzing the strongly coupled, supersymmetric
gauge theories.  Moreover, the proposed large $N$ dual provides a simpler
and more unified approach for obtaining exact results
for this class of supersymmetric gauge theories.

\Date{March 2001}

\def\Lco{\Lambda_{0}}
\def\Dd{\Delta_{21}}
\def\Dc{\Delta_{43}}
\def\Dph{\Delta}
\def\Tpi{{1\over 2\pi i}}

\def\np#1#2#3{Nucl. Phys. B {#1} (#2) #3}
\def\pl#1#2#3{Phys. Lett. B {#1} (#2) #3}

\def\prl#1#2#3{Phys. Rev. Lett. {#1} (#2) #3}
\def\physrev#1#2#3{Phys. Rev. D {#1} (#2) #3}

\def\N{{\cal N}}
\def\ev#1{\langle#1\rangle}

\lref\GubKleb{S. Gubser and I. Klebanov, ``Baryons and Domain Walls in an N=1 Superconformal 
Gauge Theory,'' hep-th/9808075, Phys. Rev. D58 (1998) 125025}
\lref\stromc{A. Strominger,``Massless Black Holes and Conifolds in String 
Theory,''
hep-th/9504090, Nucl.Phys. B451 (1995) 96.}
\lref\katmor{S. Katz and D. Morrison, ``Gorenstein Threefold Singularities
with Small Resolutions Via Invariant Theory For Weyl Groups,''
J. Algebraic Geometry 1 (1992) 449-530}
\lref\katkach{S. Kachru, S. Katz, A. Lawrence, J. McGreevy, ``Open string
instantons and superpotentials,'' hep-th/9912151, Phys. Rev. D62 (2000) 026001}
\lref\ILS{K. Intriligator, R.G. Leigh, and N. Seiberg, ``Exact Superpotentials
in Four Dimensions,'' hep-th/9403198, \physrev{50}{1994}{1092}.}
\lref\Iin{K. Intriligator, ```Integrating in' and exact superpotentials in 
4d,'' hep-th/9407106, \pl{336}{1994}{409}.}
\lref\nati{N. Seiberg, ``Naturalness Versus Supersymmetric Non-renormalization
Theorems,'' hep-ph/9309335, \pl{318}{1993}{469}.}
\lref\SW{N. Seiberg and E. Witten, ``Monopole Condensation and Confinement in
N=2 Supersymmetric Yang-Mills Theory,'' hep-th/9407087, \np{426}{1994}{19}.}
\lref\AF{P.C. Argyres and A.E. Faraggi, ``The Vacuum Structure and Spectrum
of N=2 Supersymmetric SU(N) Gauge Theory,'' hep-th/9411057,
\prl{74}{1995}{3931}.}
\lref\KLTY{A. Klemm, W. Lerche, 
S. Theisen, S. Yankielowicz, ``Simple Singularities and N=2
Supersymmetric Yang-Mills Theory,'' hep-th/9411048,
\pl{344}{1995}{169}.}
\lref\DS{M. Douglas and S. Shenker, ``Dynamics of SU(N) Supersymmetric Gauge
Theory,'' hep-th/9503163, \np{447}{1995}{271}.}
\lref\EFGIR{S. Elitzur, A. Forge, A. Giveon, K. Intriligator, E. Rabinovici,
``Massless Monopoles via Confining Phase Superpotentials,'' hep-th/9603051,
\pl{379}{1996}{121}.}
\lref\gova{R. Gopakumar and C. Vafa, ``On the Gauge
 Theory/Geometry Correspondence,'' 
 hep-th/9811131.}
\lref\vaug{C. Vafa, ``Superstrings and Topological Strings at Large N,''
 hep-th/0008142.}
\lref\amv{M. Atiyah, J. Maldacena
and C. Vafa, ``An M-theory Flop as a Large N Duality,''
Mhep-th/0011256.}
\lref\achar{B.S. Acharya,``On Realising N=1 Super Yang-Mills in M theory,''
 hep-th/0011089.}
\lref\kutetal{D. Kutasov, ``A Comment on Duality in N=1 Supersymmetric 
Non-Abelian Gauge Theories,'' hep-th/9503086, \pl{351}{1995}{230};
D. Kutasov and A.  Schwimmer, ``On Duality in Supersymmetric
Yang-Mills Theory,'' hep-th/9505004,
\pl{354}{1995}{315}; D. Kutasov, A. Schwimmer, and N. Seiberg, ``Chiral Rings,
Singularity Theory, and Electric-Magnetic Duality,'' hep-th/9510222,
\np{459}{1996}{455}.}
\lref\st{A. Klemm, W. Lerche, P. Mayr, C.Vafa
and N. Warner, ``Self-Dual Strings and N=2 Supersymmetric Field Theory,''
hep-th/9604034.}\lref\gvw{
S. Gukov, C. Vafa and E. Witten, ``CFT's From Calabi-Yau Four-folds,''
hep-th/9906070, Nucl.Phys. B584 (2000) 69.}\lref\sv{A. Shapere
and C. Vafa, ``BPS Structure of Argyres-Douglas Superconformal Theories,''
hep-th/9910182.}
\def\dualref{\refs{\gova, \vaug}}
\lref\ArgDoug{P.C. Argyres and M.R. Douglas, ``New Phenomena in $SU(3)$ 
Supersymmetric Gauge Theory,'' hep-th/9595062, \np{448}{1995}{93};
P.C. Argyres, M.R. Plesser, N. Seiberg, and E. Witten, ``New N=2 
Superconformal Field Theories,'' hep-th/9511154, \np{461}{1996}{71}.}
\lref\witadscft{E. Witten,
``Baryons And Branes In Anti de Sitter Space,'' hep-th/9805112, JHEP
9807 (1998) 006.}\lref\grog{D.J. Gross and H. Ooguri, ``Aspects of
Large N Gauge Theory Dynamics as Seen by String Theory hep-th/9805129,
Phys.Rev. D58 (1998) 106002.}
\lref\JBYO{J. de Boer, Y. Oz, ``Monopole Condensation and Confining 
Phase of $N=1$ Gauge Theories via M Theory Fivebrane,'' hep-th/9708044,
\np{511}{1998}{155}.}
\lref\whitham{A. Gorsky, I. Krichever, A. Marshakov, A. Mironov,
A. Morozov, ``Integrability and Seiberg Witten Exact Solution,''
hep-th/9505035, \pl{355}{1995}{466}; E. Martinec and N. Warner,
``Integrable systems and supersymmetric gauge theory,'' hep-th/9509161,
\np{549}{1996}{97}; T. Nakatsu and K. Takasaki, ``
Whitham-Toda hierarchy and N = 2 supersymmetric Yang-Mills theory,''
hep-th/9505162, Mod. Phys. Lett. A11 (1996) 157;
J.D. Edelstein, M. Marino, and J. Mas, ``Whitham Hierarchies,
Instanton Corrections, and Soft Supersymmetry Breaking in $N=2$ $SU(N)$
Super-Yang-Mills Theory,'' hep-th/9805172, \np{541}{1999}{671}.}
\lref\wittens{E. Witten, ``Solutions Of Four-Dimensional Field Theories 
Via M Theory,'' hep-th/9703166, \np{500}{1997}{3}; K. Hori,
H. Ooguri, Y. Oz, ``Strong Coupling Dynamics of Four-Dimensional 
N=1 Gauge Theories from M Theory Fivebrane,'' hep-th/9706082, 
Adv. Theor. Math. Phys.
1 (1998) 1; E. Witten, ``Branes And The Dynamics Of QCD,''
hep-th/9706109, \np{507}{1997}{658}.}

\newsec{Introduction}

Partially wrapping D-branes over non-trivial cycles of non-compact
geometries yields large classes of interesting gauge theories,
depending on the choice of geometry.  It has also been suggested in
\dualref\ that $N\gg 1$ D-branes, wrapped over cycles, have a dual
description (in a suitable regime of parameters) involving transitions
in geometry, where the D-branes have disappeared and have been
replaced by fluxes.  This duality can be reformulated and explained
as a geometric flop in
the context of M-theory propagating on $G_2$ holonomy manifolds
\refs{\amv, \achar}.  In this paper, we use these ideas to
propose a new class of dualities.

The simplest case, which will be the main focus of this paper, corresponds
to an $\N=1$ supersymmetric gauge theory with adjoint chiral superfield
$\Phi$ and tree-level superpotential
\eqn\wtreei{W_{tree}=\sum _{p=1}^{n+1}{g_p\over p}\Tr\ 
\Phi ^p\equiv \sum _{p=1}^{n+1}g_pu_p,}
where the gauge group can be either $SU(N)$ or $U(N)$, depending on
whether or not we treat $g_1$ as a Lagrange multiplier imposing
tracelessness of $\Phi$.  For simplicity, we generally refer to
$U(N)$, with the understanding that the $SU(N)$ can be obtained by
imposing the Lagrange multiplier condition. Without the superpotential
\wtreei, the theory would be ${\cal N}=2$ super-Yang-Mills. The theory
with superpotential
\wtreei\ arises
\katkach\ by wrapping $N$ type IIB D5 branes on special
cycles of certain Calabi-Yau geometries; the choice of $n$ and the
parameters $g_p$ are given by the geometry.  Using the corresponding
geometric transition, we construct a dual theory without the D-branes,
but with suitable fluxes.  There is also a mirror IIA description,
involving D6 branes wrapping 3 cycles.  The IIB description
is simpler, in that there are no worldsheet instanton corrections to
the superpotential.  However, the IIA perspective is useful for {\it
explaining} the origin of these dualities, as they are related to
geometric flop transitions in M-theory on $G_2$ holonomy geometries 
\amv .

The classical theory with superpotential \wtreei\ has many vacua, where
the eigenvalues of $\Phi$ are various roots $a_i$ of 
\eqn\Wprimei{W'(x)=\sum _{p=0}^ng_{p+1}x^p\equiv g_{n+1}\prod _{i=1}^n(x-a_i).}
In the vacuum where classically $P(x)\equiv \det (x-\Phi)=\prod
_{i=1}^n (x-a_i)^{N_i}$, the gauge group is broken as
\eqn\higgsi{U(N)\rightarrow \prod _{i=1}^nU(N_i)\qquad \hbox{with}\qquad 
\sum_{i=1}^nN_i=N.}
In the geometric construction \katkach, this is seen because we can
wrap $N_i$ D5 branes on any of $n$ choices of $S^2\cong {\bf
P}^1$. Such a vacuum exists for any partition of $N=\sum _{i=1}^n
N_i$.

Applying the proposal of \dualref\ to each $S^2$, a transition occurs
where we are instead left with $n$ $S^3$s.  As we discuss, the
non-compact Calabi-Yau geometry is now given by the following surface
in ${\bf C}^4$:
\eqn\Wgeom{W'(x)^2+f_{n-1}(x)+y^2+z^2+v^2=0,}
with $W'(x)$ the degree $n$ polynomial \Wprimei\ and $f_{n-1}(x)$ a
degree $n-1$ polynomial.  As for any Calabi-Yau, we can form an
integral basis of 3-cycles, $A_i$ and $B_i$, which form a symplectic
pairing
\eqn\pairing{(A_i,B_j)=-(B_j,A_i)=\delta_{ij}, \qquad (A_i,A_j)=(B_i,B_j)=0,}
with the periods of the Calabi-Yau given by the integral of the
holomorphic 3-form $\Omega$ over these cycles.  In the present case
\Wgeom, we have $i=1\dots n$, with the $A_i$ cycles compact and the
$B_i$ non-compact.  We denote the periods as
\eqn\omper{\int_{A_i}\Omega \equiv S_i, \qquad \int_{B_i}^{\Lambda_{0}} 
\Omega \equiv 
\Pi_i = {\partial{\cal F}\over \partial S_i}}
with ${\cal F}(S_i)$ the prepotential. $\Lco$ is a cutoff needed to 
regulate the divergent $B_i$ integrals; this is actually an infrared cutoff 
in the geometry integral, which will naturally be identified with the
ultraviolet cutoff of the 4d
QFT. Using \omper, the polynomial $f_{n-1}(x)$ 
in \Wgeom\ is to be solved for in terms of the $n$ $A_i$ periods $S_i$.

\lref\tv{T.R. Taylor, C. Vafa, ``RR Flux on Calabi-Yau and Partial
 Supersymmetry Breaking,''
hep-th/9912152, Phys.Lett. B474 (2000) 130.}
\lref\may{P. Mayr, ``On Supersymmetry Breaking in String Theory 
      and its Realization in Brane Worlds,''
hep-th/0003198, Nucl.Phys. B593 (2001) 99.}

As in \vaug , the dual theory obtained after the transitions 
to the geometry \Wgeom\ has a superpotential due to fluxes through the
3-cycles of \Wgeom:
\eqn\Wdual{-\Tpi W_{eff}=\sum _{i=1}^n(N_i\Pi _i+\alpha _i S_i),}
with $N_i$  3-form $(H_R+\tau H_{NS})$ flux through $A_i$ and $\alpha
_i$ 3-form flux $(H_R+\tau H_{NS})$
through $B_i$ \refs{\tv ,\may}.  If not for the superpotential
\Wdual, the dual theory would yield a 4d, ${\cal N}=2$ supersymmetric,
$U(1)^n$ gauge theory, with the $S_i$ the ${\cal N}=1$ chiral
superfields in the ${\cal N}=2$ $U(1)^n$ vector multiplets.  In terms
of this field theory, the superpotential \Wdual\ corresponds to
breaking ${\cal N}=2$ to ${\cal N}=1$ by adding electric and magnetic
Fayet-Iliopoulous terms \ref\fiterm{I. Antoniadis, H. Partouche,
T.R. Taylor, ``Spontaneous Breaking of N=2 Global Supersymmetry,''
hep-th/9512006, Phys.Lett. B372 (1996) 83 \semi S. Ferrara,
L. Girardello, M. Porrati, ``Spontaneous Breaking of N=2 to N=1 in
Rigid and Local Supersymmetric Theories hep-th/9512180,
Phys.Lett. B376 (1996) 275 \semi H. Partouche, B. Pioline, ``Partial
Spontaneous Breaking of Global Supersymmetry,'' hep-th/9702115,
Nucl.Phys.Proc.Suppl. 56B (1997) 322. }.  There will be $\N =1$
supersymmetric vacua, with the $S_i$ massive and thus fixed to some
particular $\ev{S_i}$, but with the ${\cal N}=1$ $U(1)^n$ gauge fields
left massless.  In the applications we consider, all $\alpha_j\sim 1/g_0^2$,
the bare gauge coupling of the gauge theory; this combines in a natural
way with $\Lambda_0$, replacing the
cutoff with the physical scale $\Lambda$ of the gauge theory.

The duality proposal, generalizing that of \vaug , is that these
$U(1)^n$ gauge fields coincide with those of the original theory
\higgsi\ after the $SU(N_i)$ get a mass gap and confine.  In
particular, the {\it exact} quantum effective gauge couplings $\tau
_{ij}(g_r, \Lambda; N_i)$ of the remaining massless $U(1)^n$ gauge
fields should be given by the 
prepotential of the above dual, $\tau _{ij}=\partial ^2{\cal
F}/\partial S_i\partial S_j$, evaluated at $\ev{S_i}$. Further, as in
\vaug, the $S_j$ are to be identified with the $SU(N_j)$ 
``glueball'' chiral superfields $S_j=-{1\over 32\pi ^2}\Tr
_{SU(N_j)}W_\alpha W^\alpha$, whose first component is the $SU(N_j)$
gaugino bilinear.  Finally, we claim that the superpotential \Wdual\
is the {\it exact} quantum effective superpotential of the low-energy
$SU(N)$ theory with superpotential \wtreei, in the vacuum with the
Higgsing \higgsi.

Note that the $U(1)^n$ dual theory only knows about the values of the
$N_i$ via the coefficients appearing in \Wdual.  In particular, the
$\Pi _i(S_j;g_r, \Lambda )$ and ${\cal F}(S_i; g_r, \Lambda)$ are
completely {\it independent} of the $N_i$, depending only on $\Lambda$ and
the parameters $g_r$ via \Wgeom.  Upon adding \Wdual\ to the dual
theory, one obtains $\ev{S_i}$ which are complicated functions of the
$N_i$, $g_r$, and $\Lambda$.  Integrating out the $S_i$ gives the
exact quantum 1PI effective superpotential $W_{eff}(g_r,\Lambda, N_i)$
of the original theory.

The geometric transition leads to a new duality, which can be stated
in purely field theory terms: the $U(N)$ theory with adjoint and 
superpotential \wtreei\ is dual to a $U(1)^n$ theory and superpotential 
\Wdual.   This duality is reminiscent of that of \kutetal.

The above duality makes some highly non-trivial predictions for the
exact $U(1)^n$ gauge couplings $\tau _{ij}(g_r, \Lambda)$ and the
exact effective superpotential $W_{eff}(g_r, \Lambda)$.  This allows
us to check the duality, by comparing with the exact results which can
(at least in principle) be obtained for these quantities via a direct
field theory analysis.  The above quantities can be exactly obtained
(again, at least in principle) by viewing the $\N=1$ $U(N)$ theory
with adjoint $\Phi$ and superpotential \wtreei\ as a deformation of
$\N=2$, and using the known exact results for $\N =2$ field theories.
We find perfect agreement between these results, which is a highly
non-trivial check of our proposed duality.

The organization of this paper is as follows: In section 2 we review
the large $N$ duality of \vaug\ for ${\cal N}=1$ Yang-Mills theory, and
briefly discuss the extension to include massive flavors in the
fundamental of $U(N)$.  In section 3, we discuss how to geometrically
engineer the general $\N=1$ theory with adjoint and superpotential
\wtreei. In section 4 we propose the large $N$ dual of these theories via
the transition in the CY geometry where $S^2$s are blown down, $S^3$s
are blown up, and the branes have been replaced with fluxes.  In
section 5 we analyze the $U(N)$ theory with adjoint and superpotential
using standard supersymmetric field theory tools.  In section $6$ we
specialize these results to the case of the cubic superpotential.  In
section 7 we analyze the proposed large $N$ duals and show how the
leading order computation of gauge theory based on gaugino condensate
follows from monodromies of the geometry.  In section 8 we specialize
to the cubic superpotential and compute exact results for the quantum
corrected superpotential using the proposed dual.  We find perfect
agreement with the results based on a direct gauge theory analysis.
In appendix A we present the details of the analysis for
one of the field theory examples, and in appendix B we discuss
the series expansion for computing the periods for the case of cubic
superpotential.

\newsec{Review of the large N duality for ${\cal N}=1$ Yang-Mills}

Consider type IIA string theory on a non-compact Calabi-Yau threefold
of $T^* S^3$, i.e. the conifold, with defining equation given by
$$x^2+y^2+z^2+v^2=\mu,$$
and consider wrapping $N$ D6 branes on the $S^3$, with the unwrapped
dimensions filling the Minkowski spacetime. This gives rise to a 4d
$\N=1$ $U(N)$ pure Yang-Mills theory.  The duality proposed in \vaug ,
which was motivated by embedding the large $N$ topological string
duality of \gova\ into superstrings, states that in the large $N$
limit this theory is equivalent to type IIA strings propagating on the
blow up of the conifold.  This is a geometry involving a rigid sphere
${\bf P}^1$, where the normal bundle to the ${\bf P}^1$ in the CY is
given by a ${\cal O}(-1) +{\cal O}(-1)$ bundle over it (i.e. two
copies of the spinor bundle over the sphere).  The branes have
disappeared and have been replaced by an RR flux through ${\bf P}^1$
and an NS flux on the dual four cycle \vaug.  This duality has been
embedded into M-theory, where it admits a purely geometric
interpretation \refs{\amv, \achar}.  The $SU(N)$ gauge theory
decouples from the bulk in the limit where the size $S$ of the blowup
sphere ${\bf P}^1$ is small.  The size $S$ is fixed in terms of the
units of flux, and the appropriate decoupling limit is large $N$.  $S$
gets identified \vaug\ with the glueball superfield $S=-{1\over 32 \pi
^2}\Tr W_\alpha W^{\alpha}$ of the $SU(N)$ theory, so its expectation value
corresponds to gaugino condensation in the $SU(N)$ theory.

As noted in \vaug\
one can also consider the mirror description of this geometry, which is
simpler to work with (as the worldsheet instanton corrections to
spacetime superpotential are absent).  This corresponds to switching
{}from IIA to IIB theory and reversing the arrow of transition: the
original $U(N)$ theory is obtained from type IIB $D5$ branes
wrapped around the ${\bf P}^1$ in the blown up conifold geometry and,
in the large $N$ limit, this is equivalent to type IIB on the
deformed conifold background:
$$f=x^2+y^2+z^2+v^2-\mu=0.$$
The deformation parameter $\mu$ will, again, be identified with the
$SU(N)$ glueball superfield.  Rather than the $N$ original D5 branes,
there are now $N$ units of RR flux through $S^3$, and also some NS
flux through the non-compact cycle dual to $S^3$.  This mirror
description is related to a particular limit of the large $N$ duality
proposed in \ref\kst{I.R. Klebanov and M.J.  Strassler, ``Supergravity
and a Confining Gauge Theory: Duality Cascades and $\chi$SB-Resolution
of Naked Singularities,'' hep-th/0007191, JHEP 0008 (2000) 052.}\ and
\ref\mn{J. M. Maldacena and C. Nunez,
``Towards the large N limit of pure N=1 super Yang Mills,''
hep-th/0008001, Phys.Rev.Lett. 86 (2001) 588.}.

The value of the modulus $\mu$  is fixed \vaug\ by the fluxes,
and this is captured by a superpotential for $S$,
whose first component is proportional to $\mu$.  Specializing \pairing\ and
\omper\ to the conifold, we have a single compact 3-cycle $A\cong S^3$, 
and a single dual, non-compact 3-cycle $B$.  The $A$ period of the
holomorphic 3-form $\Omega$ is $S$. There are $N$ units of RR flux
through $A$, and the NS flux $\alpha$ through $B$; $\alpha$ is
identified with the bare coupling of the 4d $U(N)$ gauge theory.
 
The holomorphic three-form $\Omega$ is given by
$$\Omega ={dxdydzdv\over df}\sim {dxdydz\over v} ={dxdydz\over
{\sqrt{\mu-x^2-y^2-z^2}}}$$
The 3-cycles $A$ and $B$ can be viewed as 2-spheres spanned by  
a real subspace of $y, z$ fibered over $x$, as in \refs{\st, \gvw ,\sv},
and integrating $\Omega$ over the fiber $y,z$ yields a one-form 
$\omega$ in the  $x$-plane:
$$\int_{S^2}\Omega\sim dx\sqrt{x^2-\mu } =\omega.$$
The A-cycle, projected to the x-plane, becomes an 
interval between $x=\pm {\sqrt \mu}$. Thus the A-period
is given by: 
$$S=\int _A\Omega =\Tpi\int_{-{\sqrt
\mu}}^{\sqrt{\mu}}dx\sqrt{x^2-\mu}={\mu\over 4}$$
The B-period can be viewed as an integral from $x=\sqrt \mu$ to
infinity.  However this integral is divergent, and thus must be
cutoff to regulate the infinity.  Giving $S$ 
dimension $3$, $x$ has dimension 3/2, so we put the cutoff
at $x=\Lambda_0^{3/2}$ where
$\Lambda_0$ has mass dimension 1:
$$\Pi =\Tpi\int_{\sqrt \mu}^{\Lambda_0^{3/2}}dx\sqrt{x^2-\mu}
=\Tpi \left({1\over 2}\Lco^3-3S\log \Lco-S(1-\log S)\right)+
{\cal O}(1/\Lambda_0)$$
Note that, under $\Lco^3\rightarrow e^{2\pi i} \Lco^3$, $\Pi
\rightarrow \Pi - S$, shifting the $B$ period by an $A$ period.
Using the fact that we have $N$ units of RR flux through $S^3$ and
$\alpha$ units of NS flux through the B-cycle, we find the
superpotential \vaug :
$$W_{eff}=N [3 S\log \Lco+S(1-\log S)]-2\pi i\alpha S.$$
$\alpha$ is related to the bare coupling constant of the $SU(N)$ gauge
theory by $2\pi i\alpha =8\pi ^2/g_0^2$.  The coefficient of $S$ in the
above superpotential is given by
$$S(3N{\rm log}\Lambda_0 -2\pi i\alpha),$$
which is the geometric analog of the running of the coupling.  $\alpha$
depends on $\Lambda_0$ in such a way that the above quantity is finite
as $\Lambda_0\rightarrow \infty$:
$${8\pi ^2\over g^2(\Lambda_0)}={\rm const.}+3N{\rm log}\Lambda_0,$$
which is exactly the expected running of the coupling constant
for the 4d ${\cal N}=1$ $U(N)$ Yang-Mills theory.
The upshot is to replace the cutoff $\Lambda_0$ in the above
expression with the scale of the gauge theory, which we will denote by
$\Lambda$.  We thus have for the superpotential
$$W_{eff}=S{\rm log}[\Lambda^{3N}/S^N] + N S$$
(the linear term $NS$ is a matter of convention and
defines what one means by the physical scale $\Lambda$).
This is indeed the superpotential of
\ref\Spot{G. Veneziano and S.  Yankielowicz, \pl{113}{1982}{321};
T.R. Taylor, G. Veneziano, and S. Yankielowicz, \np{218}{1983}{493}.}
for the massive glueball $S$. Integrating out $S$ via $dW_{eff}/dS=0$ leads
to the $N$ supersymmetric vacua of $SU(N)$ ${\cal N}=1$ supersymmetric
Yang-Mills:
$$\ev{S}=e^{2\pi i k/N}\Lambda ^3, \qquad k=1, \dots N.$$

\subsec{Gauge Theoretic Reformulation of the duality}

We can formulate the above large $N$ duality in purely gauge theoretic
terms. The conifold geometry without the fluxes corresponds to an
${\cal N}=2$ $U(1)$ gauge theory with a charged hypermultiplet
\stromc.  Turning on fluxes is equivalent to adding
electric and magnetic Fayet-Iliopoulous superpotential terms, which
softly break $\N=2$ to ${\cal N}=1$.  The $\N=2$ vector multiplet
consists of a neutral $\N=1$ chiral superfield $S$ and an $\N=1$
photon.  The $\N=1$ $U(1)$ photon is
left massless, and is to be identified with the overall $U(1) \subset
U(N)$ of the original ${\cal N}=1$ theory.  The $\N=1$ chiral superfield
$S$ gets a mass, and is to be identified with the massive glueball
chiral superfield $S$ of the $SU(N)$ theory.

The identification of the $U(1)$ of the dual theory with the
$U(1)\subset U(N)$ is consistent with the fact that minimization of
the superpotential gives rise to
$$N{\partial \Pi\over \partial S}+\alpha =N \tau +\alpha=0$$
where we used the special geometry to connect the periods
of the B-cycles with the coupling constant $\tau $
of the $U(1)$.  Note
that the coupling of the $U(1)$ theory is $-\alpha/N$ as it should be
where $-\alpha $ is the bare coupling of the $U(N)$ theory and
$U(1)$ is identified with $1/N$ times the identity matrix
in $U(N)$ adjoint.  In fact the ``charged hypermultiplet''
of the $U(1)$ is nothing but the baryon field of the original
$U(N)$ theory.  To see this note that before turning on the
RR flux on $S^3$, wrapping a $D3$ brane around it gives
a charged hypermultiplet. Turning on the RR flux, induces $N$
units of fundamental charge on it, as noted in the context
of AdS/CFT correspondence in \refs{\witadscft, \grog, \GubKleb}.
After turning on the flux the field is not allowed by itself,
i.e., it is attached to $N$ fundamental strings going off to infinity.
Thus after the FI deformations
of the superpotential it is slightly misleading to think of the $U(1)$
theory as having a fundamental hypermultiplet.  In that context
one can simply view this as an effective $U(1)$ theory with the 
SW ${\cal N}=2$
geometry as would have been the case with a fundamental hypermultiplet.

\subsec{Adding Massive Fields}

As discussed in \vaug, we can also consider adding some quark chiral
superfields, in the fundamental representation of $SU(N)$.  In the type
IIB description this is done by taking a D5 brane wrapping a
holomorphic 2-cycle not intersecting the ${\bf P}^1$, but separated by
a distance $\rho$, where $\rho$ is proportional to the mass of the
hypermultiplet, as the matter comes from strings stretching between
the non-compact brane and the $N$ branes wrapped on ${\bf P}^1$.  If
$(\zeta_1,\zeta_2)$ denote the ${\cal O}(-1)+{\cal O}(-1)$ bundle over
${\bf P}^1$, the 2-cycle is the curve $(\zeta_1,\zeta_2)=(\rho ,0)$
over a point on ${\bf P}^1$.  Passing this through the conifold
transition, which in these coordinates is given by
$$\zeta_1 a-\zeta_2 b=\mu,$$
and rewriting it by a change of variables in the form
$$F(x,y)=x^2+y^2-\mu =\zeta_2 b,$$
we have a D5 brane wrapping a 2-cycle given by $\zeta_2 =0$ and
$x=\rho$.  Since here $x$ has dimension $3/2$, and $\rho$ should be
proportional to the mass $m_0$, we identify $\rho =m_0
\Lambda_0^{1\over 2}$.
As discussed in \ref\agv{M. Aganagic and C. Vafa,
``Mirror Symmetry, D-Branes and Counting Holomorphic Discs,''
hep-th/0012041 .}\ such a D-brane gives rise
to an additional spacetime superpotential 
$$\Delta W_{eff} =
\half\int_{m_0\Lambda_0^{1/2}}^{\Lambda_0^{3/2}} 
 dx\sqrt{x^2-\mu}=
 S\log \left({m_0\over \Lco}\right)+{\cal O}({1\over \Lambda_0}).$$
This gives the running of the mass parameter with the cutoff
$\Lambda_0$.  We define the renormalized mass by $m/\Lambda
=m_0/\Lambda_0$.  Generalizing to any number of matter fields in the
fundamental representation, with mass matrix $m$, we find
$$\eqalign{W_{eff}&=S{\rm log}[\Lambda^{3N}/S^N] + NS + S 
{\rm Tr}{\rm log} [m/\Lambda ]\cr
&=S[{\rm log}{\Lambda^{3N-N_f}{\rm det}\ m \over S^N} + N].}$$
Integrating out $S$ via $dW_{eff}/dS=0$ yields the correct field theory
result: 
$$S^N=\Lambda^{3N-N_f}{\rm det} \ m.$$

\bigskip
\centerline{\epsfxsize=0.85\hsize\epsfbox{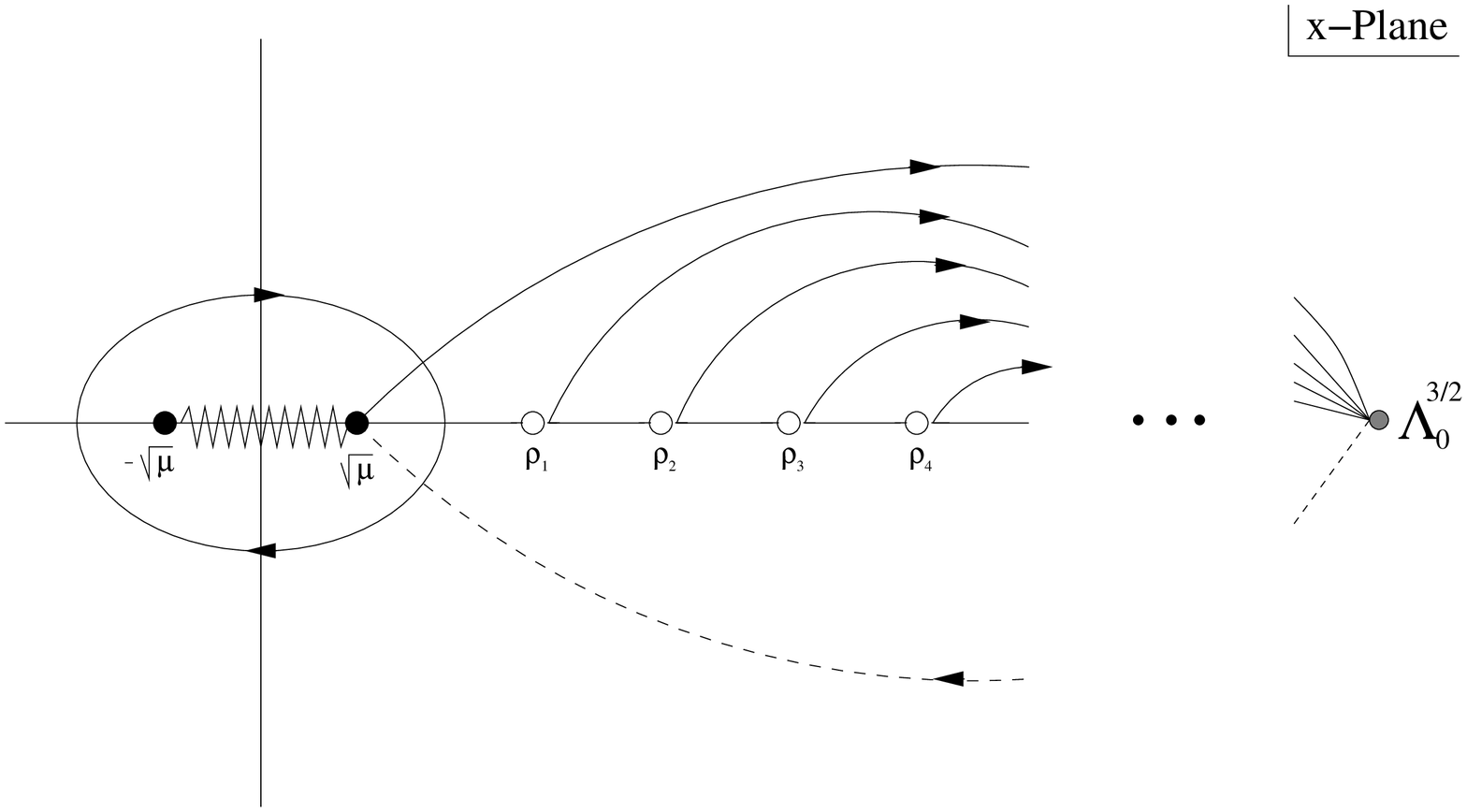}}
\noindent{\ninepoint\sl \baselineskip=8pt {\bf Figure 1}: {\sl Location of
the branch cut in the x-plane. Contours of integration of the different
periods of the geometry including those coming from massive fields.}}
\bigskip

\newsec{Geometric engineering ${\cal N}$=1 theories with adjoint $\Phi$
and superpotential $W_{tree}(\Phi)$}

The $\N=1$ $SU(N)$ Yang-Mills theory of the previous section can be
regarded as a special case of the more general theory with adjoint
$\Phi$ and superpotential as in \wtreei,
\eqn\wtree{W_{tree}(\Phi)=\sum_{p=1}^{n+1} {g_p\over p} \Tr \ \Phi^{p}.}
For $n=1$, the adjoint gets a mass $m=g_2$ and we recover the case
reviewed in the previous section.  We here review the geometric
construction of \katkach\ for general $n$.

For $W_{tree}(\Phi)=0$, the 4d field theory would be pure 
${\cal N}=2$ Yang-Mills system.  To geometrically
engineer that, all we need is a ${\bf P}^1$ in a Calabi-Yau manifold
for which the normal bundle is ${\cal O}(-2)+{\cal O}(0)$ (i.e.
it has the same normal geometry as if the ${\bf P}^1$ were in a $K3$).
If we wrap $N$ D5 branes around the ${\bf P}^1$ we obtain an ${\cal N}=2$
$U(N)$ gauge theory in the uncompactified worldvolume of the D5 brane.
The adjoint scalar $\Phi$ gets identified with the deformations
of the brane in the ${\cal O}(0)$ direction, normal to the ${\bf P}^1$.

To describe the geometry in more detail, let $z$ denote the coordinate
in the north patch of ${\bf P}^1$ and $z'=1/z$ in the south patch.
Let $x,x'$ denote the coordinate of ${\cal O}(0)$ direction in the
north and south patches respectively, and let $u,u'$ denote the
coordinates of ${\cal O}(-2)$ in the north and south patches
respectively.  Then we have
\eqn\Niigeom{z'=1/z, \qquad x'=x, \qquad u'=uz^2.}
There is a continuous family of ${\bf P}^1$s, labeled by arbitrary
$x$, at $u=0=u'$.  Each of the $N$ D5 branes can wrap a ${\bf P}^1$ at
any value of $x$.  In the $\N=2$ gauge theory living in the
unwrapped directions, this freedom to choose any $x$ for each brane
corresponds to moving along the Coulomb branch, with the $a_i$ of each
brane corresponding to an eigenvalue of the adjoint field $\Phi$.

This connection between $x$ and the Coulomb branch moduli makes it clear how
the geometry must be deformed to obtain the $\N =1$ theory with superpotential
\wtree.  Rather than having the ${\bf P^1}$, with coordinate $z$ and $z'$
at the point $u=u'=0$, for arbitrary $x$, it should exist only for
particular values of $x$, namely the values $x=a_i$ where
$W'(x)\equiv g_n\prod _{i=1}^n(x-a_i)=0$.  This is the case if \Niigeom\ 
is deformed to 
\eqn\Nigeom{z'=1/z, \quad x'=x, \quad u'=uz^2+W'(x)z,}
which is indeed only compatible with $u=u'=0$ at the $n$ choices of $x=a_i$
where $W'(x)=0$.  Note that now we can distribute the $N$ D5-branes among
the vacua $a_i$, i.e. $N_i$ branes wrapping the corresponding $S^2$
at $x=a_i$.
This gives a geometric realization of the breaking of $U(N)\rightarrow
\prod_i U(N_i)$.

\newsec{Large N Duality Proposal}

We now obtain the large $N$ dual of the $U(N)$ theory with adjoint
$\Phi$ and superpotential $W_{tree}(\Phi)$ by considering the geometric
transition where each of the $n$ ${\bf P}^1$'s have shrunk and have
been replaced by a finite size $S^3$.  As already mentioned, the
sizes of the $n$ $S^3$s will correspond to the non-zero gaugino
condensation expectation values in the $n$ factors of ${\cal N}=1$
non-Abelian gauge groups in \higgsi.  The needed blow-down of the $n$
${\bf P^1}s$ of the geometry of the previous section has been discussed
in \katmor\ and we will review it here. 
We start with the defining equation \Nigeom.  Its blowdown 
can be obtained by the change of variables as follows: define
$x_1\equiv x$, $x_2\equiv u'$, $x_3\equiv z'u'$, $x_4\equiv u$; 
using \Nigeom, these satisfy
$$x_2x_4-x_3^2+x_3 W'(x_1)=0.$$
By completing the square involving $x_3$ and $W'$ and redefining
the variables slightly we obtain the equation
\eqn\fineq{W'(x)^2+y^2+z^2+v^2=0.}
This geometry is singular, even for a generic $W'(x)$; near each
critical point of $W(x)$ it has the standard conifold singularity.
The large $N$ dual follows from desingularizing the geometry
\fineq, allowing the $n$ $S^3$s to have
finite size, rather than zero size as in \fineq.

\subsec{Desingularization of the Geometry}

Consider the most general desingularization of \fineq, subject to the
restriction of \gvw\ that the deformation be a normalizable mode.  For
the case at hand, as $W'^2$ is a polynomial of degree $2n$, the most
general desingularization of
\fineq\ subject to the normalizability restriction is to add a
polynomial $f_{n-1}(x)$ of degree $n-1$ in $x$
\sv , giving the geometry
\eqn\geomb{W'(x)^2+f_{n-1}(x)+y^2+z^2+v^2=0.}
Under this deformation, each of the $n$ critical points $x=a_i$
\Wprimei\ (where $W'=0$) splits into two, which we denote as $a_i^+$
and $a_i^-$. 

As in the case of the conifold, the period integrals of the
holomorphic three-form over the $A_i$ and $B_i$ cycles can be written
as integrals of an effective one-form $\omega $ over projections of the
cycles to the $x$ plane.   As in the conifold case, the non-trivial
3-cycles have simple projections to the $x$ plane.
The one-form $\omega$ is given by doing the $\Omega$
integral over the fiber $S^2$ cycles (corresponding to the $y,z,v$ 
coordinates on the surface \Nigeom); this gives
\eqn\effone{\omega =dx\sqrt{W'^2(x)+f_{n-1}(x)}.}
Therefore, the periods of the holomorphic three-form $\Omega$ over
the $n$ 3-cycles $A_i$ of \geomb, which are compact 3-spheres, are given by,
\eqn\fperiod{S_i= \pm \Tpi\int_{a_i^-}^{a_i^+}\omega}
where the sign depends on the orientation; the periods over the dual
$B_i$ cycles are
\eqn\dualperiod{\Pi_i=\Tpi\int_{a_i^+}^{\Lco} \omega.}

The map between the $n$ coefficients in $f_{n-1}(x)$
and the $S_i$ can thus be obtained by direct computation, and
$f_{n-1}(x)$ can then be solved for as particular functions
$f_{n-1}(x; S_i)$.


\bigskip
\centerline{\epsfxsize=1.05\hsize\epsfbox{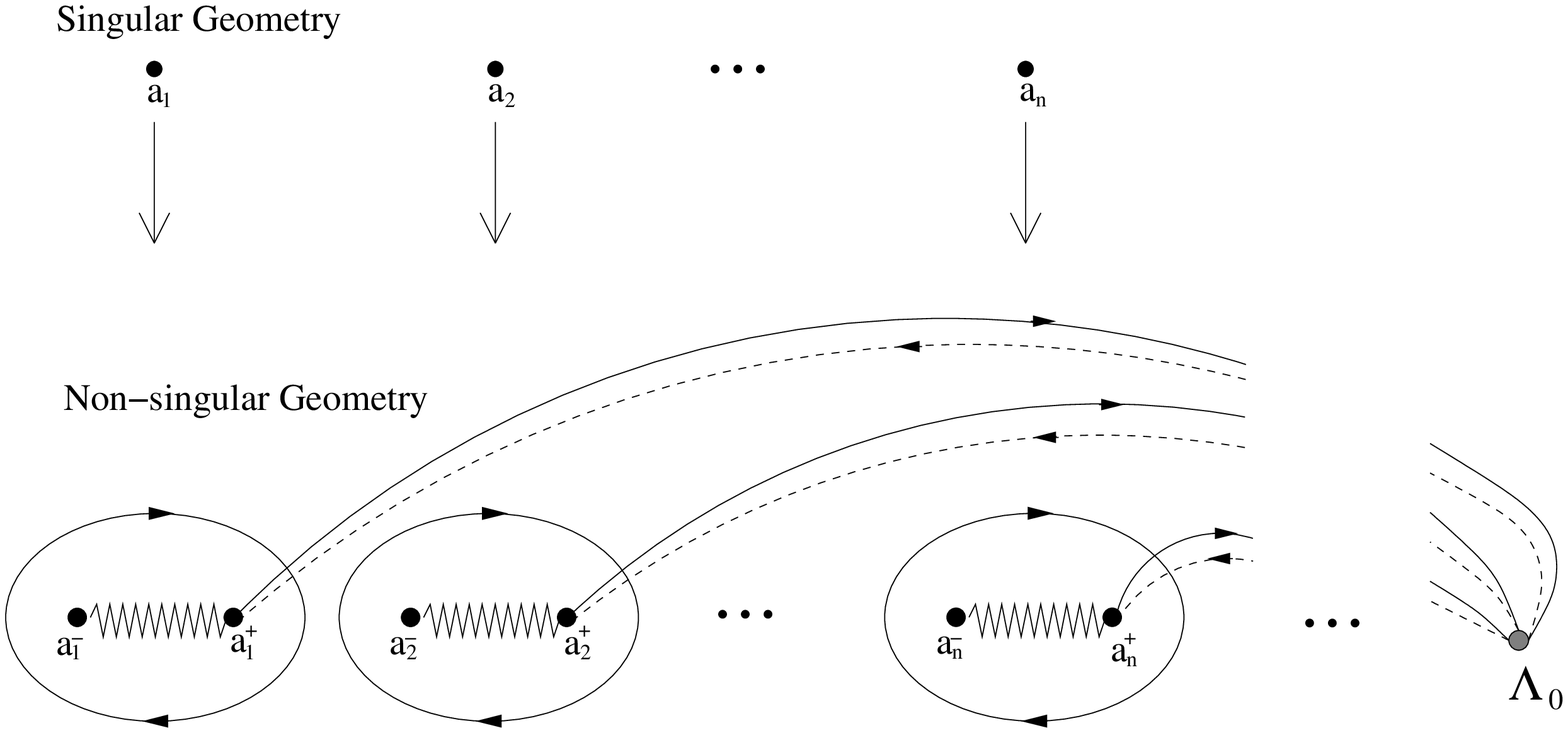}}
\noindent{\ninepoint\sl \baselineskip=8pt {\bf Figure 2}: {\sl Geometry
before and after introducing the deformation $f_{n-1}(x)$. The choice of
branch cuts and integration contours for the different periods is also
shown. Dashed lines are paths on the lower sheet.}}
\bigskip


As we already mentioned, the $n$ values of $S_i$ are mapped
under the duality to the $n$ glueball fields $S_i=-{1\over 32 \pi ^2}
\Tr _{SU(N_i)}W_{\alpha}W^{\alpha}$ for the non-Abelian factors in \higgsi.
(The $S_i$ can be defined in a gauge invariant way.)  Just as with the
case of pure $\N =1$ $U(N)$ Yang-Mills, the $S_i$ of the dual theory
will become massive and obtain particular expectation values thanks to
a superpotential $W_{eff}$, with the expectation values
$\ev{S_i}$ determined from finding the critical points of $
W_{eff}$.  The dual superpotential $W_{eff}$ arises from the 
non-zero fluxes left after the transition.

Rather than having D-branes, as present before the transition, the
above deformed geometry will have $N_i$ units of $H_R$ flux through
the $i$-th $S^3$ cycle $A_i$.  In addition, there is an $H_{NS}$ flux
$\alpha$ through each of the dual non-compact $B_i$ cycles, with
$2\pi i\alpha = 8\pi ^2/g_0^2$ given in terms of the bare coupling constant
$g_0$ of the original 4d $U(N)$ field theory.  We thus have the
superpotential, given in terms of the $A_i$ and $B_i$ periods \omper\
as
\eqn\suppo{-\Tpi W_{eff}=\sum_{i=1}^n N_i \Pi_i+
\alpha (\sum_{i=1}^{n} S_i).}
This $W_{eff}$ depends on the coefficients $g_r$ of the classical
superpotential \wtreei\ of the original $U(N)$ theory with adjoint by
way of the geometry \geomb.  $ W_{eff}$ is a function of the $n$
$S_i$, or equivalently the $n$ unknown parameters in $f_{n-1}(x)$.
The supersymmetric vacua have fixed $\ev{S_i}$, obtained by solving
\eqn\Sieom{{\partial W_{eff}\over \partial S_i}=0, \qquad i=1\dots n.}
These $\ev{S_i}$ will depend on the $N_i$, the parameters $g_r$ entering
in the original $W_{tree}(\Phi)$ and thus on the geometry \geomb,
and $\Lambda_0$, the $B_i$ integral infrared cutoff.

In the classical limit, where we set the $S_i$ to zero,
and thus $f_{n-1}(x)=0$, the period of the one-form \effone\ gives
\eqn\classlim{\Pi_i=\Tpi\int_{a_i}^{\Lambda_0} dx W'(x)=
\Tpi\left( W(\Lambda_0)-W(a_i)\right).}
Then the dual superpotential is $W_{eff}= \sum _i N_iW(a_i)$ 
(ignoring the irrelevant constant $W(\Lambda_0)$).  This indeed matches
with the classical superpotential of the original $U(N)$ theory, given
by simply evaluating the superpotential \wtreei\ in the vacuum with breaking
\higgsi, where $N_i$ eigenvalues of the $\Phi$ field
take eigenvalue $a_i$.

\subsec{Aspects of the $U(1)^{n}$ gauge fields}

The dual theory obtained after the transition is an $\N =2$ $U(1)^n$
gauge theory, broken to $\N =1$ $U(1)^n$ by the superpotential $
W_{eff}$ \suppo. The $S_i$, which are in the same $\N =2$ multiplet as
the $U(1)^n$, get masses and frozen to particular $\ev{S_i}$ by
$W_{eff}$.  On the other hand, the $\N =1$ $U(1)^n$ gauge fields
remain massless. The couplings $\tau _{ij}$ of these $U(1)$'s can be
determined from $\Pi _i(S)$ or the ${\cal N}=2$ prepotential ${\cal F}(S_i)$,
with $\Pi_i =\partial {\cal F}/ \partial S_i$, of the geometry under
consideration:
\eqn\tauis{\tau_{ij}={\partial \Pi_i 
\over \partial S_j}={\partial ^2 {\cal F}(S_i)\over \partial S_i\partial S_j}.}
The couplings
\tauis\ should be evaluated at the $\ev{S_i}$ obtained from \Sieom. 

Note that \Sieom\ and \tauis\ imply
\eqn\taudc{\sum_i N_i \tau_{ij}+\alpha =0.}
We identify the $F_i$ the i-th block $U(1)$ field strength
with the generator in $U(N)$ which is $1/N_i$ times the identity 
matrix in the $i$-th block and zero elsewhere. In this way the
$F_i-F_j$ correspond to field strengths of the $U(1)^{n-1}$'s
coming from the $SU(N)$ and the $N_i F_i$ will corresponds
to the overall $U(1)$.  Thus the above equation is
 consistent with the fact that the overall $U(1)$ is a linear
combination of the $U(1)^n$'s with coefficients given by $N_i$,
together with the fact that the bare coupling constant of the
overall $U(1)$ should be the same as that of the original $U(N)$
theory, as the $U(1)$ is decoupled.  Moreover it is consistent
with the fact that there is no coupling between the field strength
of this overall $U(1)$ with the other $U(1)^{n-1}$.  Thus extremizing
the superpotential is equivalent to this structure for the gauge
coupling constants of the $U(1)$ factors.

One can also relate the coupling constants of the $U(1)$ factors
to the period matrix of the hyperelliptic curve
$$y^2=W'(x)^2+f_{n-1}(x)$$
To see that, from \tauis\ we will have to compute the period
integrals of $\partial \omega /\partial (S_i-S_j)$ about
the cycles of the hyperelliptic curve, where
$\omega =ydx$.  As we will discuss in section 7 the coefficient
of $x^{n-1}$ of $f_{n-1}(x)$ is proportional to the sum
of $S_i$'s and thus considering $\partial \omega /\partial (S_i-S_j)$
gives rise to a linear combination of
$${x^{n-r}dx\over y}$$
with $2\le r \le n$, a basis of the $n-1$ holomorphic one-forms on the
hyperelliptic curve.  Thus $\tau_{ij}$ can be identified with the
period matrix of the hyperelliptic curve.

\subsec{Gauge theoretic reformulation}

Just as in the case of $n=1$ we can reformulate this duality in
terms of a duality of two gauge systems: We start with ${\cal N}=2$
pure Yang-Mills theory for gauge group $U(N)$ and deform it by the
superpotential $W_{tree}(\Phi )$ of degree $n+1$ in the scalar field,
breaking the $U(N)$ into $n$ factors $U(N_i)$.  The $SU(N_i)$ gaugino
bilinear together with the $U(1)\subset U(N_i)$ forms an ${\cal N}=2$
multiplet.  One considers a dual ${\cal N}=2$ multiplet containing $
U(1)^n$ softly broken to ${\cal N}=1$ by a superpotential term.  Note
that the ${\cal N}=2$ we have proposed is of the form that appears in
an ${\cal N}=2$ theory with a $U(n)$ gauge group with some matter
fields (whose structure is dictated by the superpotential).  In fact
the dual ${\cal N}=2$ system we have been considering is of the type
studied in \ArgDoug\ and was connected to a type IIB description
considered here in \sv .  In such a formulation the decoupling of the
overall $U(1)$ from the other $U(1)$'s occurs as in \taudc, consistent
with the minimization of the superpotential.

\newsec{Field theory analysis}

We now analyze the strong coupling dynamics of the $U(N)$ theory, with
adjoint $\Phi$ and superpotential \wtreei, in the vacuum with the
classical breaking \higgsi.  In the quantum theory, each $\N =1$ super
Yang-Mills $SU(N_i)$ in \higgsi\ generally confines, with $N_i$
supersymmetric vacua.  The $N_i$ vacua correspond to $N_i$-th roots of
unity phases of the gaugino condensate $\ev{S_i}\neq 0$, with
$S_i=-{1\over 32\pi ^2}\Tr W_{\alpha}W^{\alpha}$ the $SU(N_i)$
glueball chiral superfield.  The $U(1)^{n}$ in \higgsi\ are free,
and therefore remain unconfined and present in the low energy
theory. The vacua can also have more interesting behavior.  For
example, in $SU(3)$ with a cubic superpotential for $\Phi$ but no
quadratic mass term, the vacuum is at the non-trivial conformal field
theory point of \ArgDoug.

The low energy theory contains an effective superpotential
$W_{eff}(g_p, \Lambda)$ which gives the chiral superfield expectation
values via \ILS\
\eqn\wder{\eqalign{{\partial W_{eff}(g_p, \Lambda)\over \partial g_p}
&=\ev{u_p}\cr {\partial W_{eff}(g_p,\Lambda)\over \partial \log
\Lambda ^{2N}}&=\ev{S}\equiv\sum _{i=1}^n\ev{S_i}.}}  
$W_{eff}$ can often be obtained exactly, thanks to its holomorphic
dependence on $g_p$ and $\Lambda$
\nati.  In the present case, we'll discuss how $W_{eff}$ can indeed,
{\it in principle}, be obtained exactly via the $\N =2$ curves
\refs{\SW, \AF, \KLTY}; in practice, however, the 
result is quite difficult to obtain.

\subsec{Approximate $W_{eff}$ via naive integrating in}
 
The effective superpotential can often be obtained exactly via
starting from the low-energy effective theory and ``integrating in''
the massive matter fields \refs{\ILS, \Iin}.  As discussed in \Iin,
for this procedure to give an exact answer, one must be able to argue
that the scale matching relations are known exactly and that a
possible additional unknown contribution $W_{\Delta}$ to the
superpotential necessarily vanishes.  Our $\N =1$ theory with adjoint
$\Phi$ and superpotential \wtreei, does not admit this kind of symmetry
and limits arguments needed to prove the naive scale matching
relations and $W_{\Delta}=0$ as exact statements.  So naive
integrating in need not give the exact answer for $W_{eff}$;
nevertheless, it is still useful here for obtaining an approximate
answer.

To illustrate how naive integrating in can fail to give the exact
answer in the theory with adjoint $\Phi$, consider the vacuum where
classically $\ev{\Phi} =0$, leaving $SU(N)$ unbroken. Such a vacuum
exists for any tree level superpotential \wtreei.  The mass of $\Phi$
in this vacuum is $W''(0)=g_2\equiv m$, independent of the other
$g_p$.  The low energy theory is $\N =1$ $SU(N)$ pure Yang-Mills and
the dynamical scale $\Lambda _L$ of this theory is related to that of
the original high energy theory by matching the running gauge coupling
at the threshold scale $m$, giving $\Lambda _L^{3N}=m^N\Lambda ^{2N}$.
The low-energy theory has $N$
vacua with gaugino condensation and low-energy superpotential
\eqn\wlowi{W_{low}=e^{2\pi i k/N}N\Lambda _L^3=e^{2\pi i k/N} Nm\Lambda ^2.}
Using \wder\ one could use this to try to find the $\ev{u_r}$ in this
vacuum, but the answer would be incorrect for $SU(N)$ with $N>3$.  The
exact answer can be found from deforming the $\N =2$ curve following \DS, as
reviewed in the next subsection.  The exact effective superpotential
is found from this to be
\eqn\wexacti{W_{exact}=N\sum _{p=1}^{[{n\over 2}]}{g_{2p}\over 2p}
\Lambda ^{2p}\pmatrix{2p\cr p}.}
The $g_2$ term coincides with \wlowi, so both give the same $\ev{u_2}$,
but \wlowi\ gives all other $\ev{u_r}=0$, whereas \wexacti\ gives
higher $\ev{u_{2p}}\sim N \Lambda ^{2p}\neq 0$.  

The terms in \wexacti\ which are missing from \wlowi\ are weighted by
$g_{2p}\Lambda ^{2p}$, which should be small as compared with the
leading term $m\Lambda ^2$.  The reason is that the higher $g_{2p}$
appear irrelevant in the original $SU(N)$ description, so their
required UV cutoff should be larger than the dynamical scale $\Lambda$
in order for the theory to be well-defined, i.e. the $g_{3+n}\Lambda
^n$ should be small.  So the lesson is that naive integrating in here
needn't give the exact answer, but it does generally give the leading
term or terms.

On the other hand, naive integrating in actually does give the {\it
exact} answer for $W_{eff}$ in the vacua where $SU(N)\rightarrow
SU(2)\times U(1)^{N-2}$ \EFGIR.  In fact, the exact curve of the
entire $\N =2$ theory can be re-derived via ``integrating in'' in the
$SU(2)\times U(1)^{N-2}$ vacua \EFGIR.

We now outline the naive integrating-in procedure for the general
vacuum \higgsi.  The low-energy $\N =1$ SYM with gauge group
\higgsi\ leads to a low-energy superpotential via gaugino condensation
in each of the decoupled, non-abelian groups:
\eqn\Wlows{W_{low}=W_{cl}(g_r)+ \sum _{i=1}^ne^{2\pi i k_i/N_i}
N_i\Lambda _i^3.}
The term $W_{cl}(g_r)$ is simply the value of the classical superpotential
\wtreei, evaluated in the classical vacuum: 
\eqn\wcl{W_{cl}=\sum _{i=1}^n N_i\sum_{p=1}^{n+1}g_p{a_i^p\over p},}
with the $a_i$ defined in \Wprimei.
As in \wder, ${\partial W_{cl}(g_r)\over \partial g_r}=\ev{u_r}_{cl}$.  

The dynamical scale $\Lambda _i$ entering in \Wlows\ is that of the
low-energy $SU(N_i)$ theory, which is related to the scale 
$\Lambda$ of the high-energy theory by matching the running gauge
coupling across two thresholds: that of the massive $SU(N)/SU(N_i)$
W-bosons, and that of the mass of the field $\Phi$ in the vacuum.  The
classical masses of the W-bosons which are charged under $SU(N_i)$ are
$m_{W_{ij}}=a_j-a_i$.  The mass of the $SU(N_i)$ adjoint $\Phi _i\in
\Phi$ is classically $m_{\Phi _i}= W''(a_i)=g_{n+1}\prod _{j\neq
i}(a_j-a_i)$.  The scale $\Lambda _i$ of the low-energy $SU(N_i)$ is
thus obtained by naive threshold matching to be
\eqn\lamin{\Lambda _i^{3N_i}=\Lambda ^{2N} m_{\Phi _i}^{N_i}
\prod _{j\neq i}m_{W_{ij}}^{-2N_j}=g_{n+1}^{N_i} \Lambda ^{2N}
\prod _{j\neq i}(a_j-a_i)^{N_i-2N_j}.}

It will be useful in what follows to also integrate in the glueball
fields $S_i$:
\eqn\WlowS{W_{low}=W_{cl}(g_r)+ \sum _{i=1}^nS_i\left(\log ({\Lambda _i^{3N_i}
\over S_i^{N_i}})+N_i\right).}
The $S_i$ are massive, with supersymmetric vacua $\ev{S_i}=\Lambda _i^{3N_i}$,
and integrating out the $S_i$ leads back to \Wlows.

The final result of naive integrating in is thus expressed in terms of the
$a_i(g_r)$ as 
\eqn\Wiin{W_{low}(g_r)=\sum _{i=1}^n \left[ N_i\sum _{p=1}^{n+1}
g_p{a_i^p\over p}+S_i\left( \log({
g_{n+1}^{N_i}\Lambda ^{2N}\prod _{j\neq
i}(a_j-a_i)^{(N_i-2N_j)}\over S_i^{N_i}})+N_i\right)\right].}
The quantum term in \Wiin, coming from $SU(N_i)$ gaugino condensation,
is to be omitted when $N_i=1$; e.g. in the case of \EFGIR, where $N_1=2$
and all other $N_i=1$.  The result \Wiin\ happens to be exact when no
$N_i>2$ but, as emphasized above, \Wiin\ is only an approximation to
the exact answer in the more general case, where some $N_i\geq 3$.

\subsec{The exact $W_{exact}(g_r)$ via deforming the $\N =2$ results}

In this subsection, we obtain the exact 1PI generating function
$W_{exact}(g_r)$ by deforming the exact solution \refs{\SW, \AF,
\KLTY}\ of the $\N =2$ theory by the $W_{tree}(\Phi)$ \wtreei.  The
large $N$ duality proposal of section 4 gives the exact superpotential
$W_{exact}(g_r;S_i)$ as
\suppo, with the glueball fields included.  (As verified in section 7,
the naive integrating in result \Wiin\ is indeed an approximation to
this exact result; generally there is an infinite series expansion of
corrections to the naive formula \Wiin.)  Upon integrating out the
massive $S_i$ from $W_{exact}(g_r,S_i)$ \suppo, one obtains 
$W_{exact}(g_r)$, which we will verify indeed agrees with the field theory
result obtained in this subsection.  Our $W_{exact}(g_r;S_i)$ 
\suppo, however, contains the additional information about the
glueball fields $S_i$.  Although the $S_i$ are massive, this additional
information about their superpotential is physical; for example $\Delta
W$ between the different $\ev{S_i}$ vacua gives the BPS tension of the
associated domain walls.  Perhaps there's also a way to exactly
integrate in the $S_i$ in the context of the deformed $\N=2$ field theory,
though this is not presently known.

The $\N =2$ theory deformed by $W_{tree}=\sum _{i=1}^{n+1}g_r u_r$
only has unbroken supersymmetry on submanifolds of the Coulomb branch,
where there are additional massless fields besides the $u_r$.  The
additional massless fields are the magnetic monopoles or dyons, which
become massless on some particular submanifolds $\ev{u_p}$ \SW.  Near
a point with $l$ massless monopoles, the superpotential is
\eqn\WSWm{W=\sum _{k=1}^l M_k(u_r)q_k\widetilde q_k+\sum _{p=1}^{n+1} g_pu_p,}
and the supersymmetric vacua are at those $\ev{u_p}$ satisfying
\eqn\WSWeom{M_k(\ev{u_p})=0 \quad
\qquad\hbox{and}\qquad 
\sum _{k=1}^l{\partial M_k(\ev{u_p}) 
\over \partial u_p}\ev{q_k\widetilde q_k}+g_p=0,} 
the first equations are for all $k=1\dots l$ and the second for all
$r=1\dots N$ (with $g_p=0$ for $p>n+1$).  The
value of the superpotential \WSWm\ in this vacuum is simply
\eqn\WSWl{W_{eff}=\sum _{p=1}^{n+1}g_p\ev{u_p},}
with $\ev{u_p}$ the solution of $M_k(\ev{u_p})=0$, where the monopoles
are massless.   The explicit monopole masses $M_k(u_r)$ on the Coulomb 
branch can be obtained via the appropriate periods of the one-form \AF, 
\eqn\lamSW{M_k=\oint _{\gamma _k}\lambda, \quad\hbox{with}\quad 
\lambda ={1\over 2\pi i}{x\over y}{\partial P_N(x)\over \partial x}dx, \quad
\hbox{satisfying}\quad {\partial \lambda\over \partial s_r}\sim {x^{N-r}dx\over
y}+d(\dots),}
but this will not be needed here.

In the vacuum \higgsi, there are $n$ massless photons, whereas 
the original $\N =2$ theory had $N$ massless photons.  So the vacuum
\higgsi\ must have $N-n$ mutually local magnetic monopoles being massless
and getting an expectation value as in \WSWeom, $\ev{q_k\widetilde q_k}
\neq 0$ for $k=1\dots N-n$.   It can indeed be shown from \WSWeom\ that
if the highest Casimir with nonzero $g_p$ in $W_{tree}$ is $u_{n+1}$,
as in \wtreei, then the supersymmetric vacuum necessarily has at least
$l=N-n$ mutually local monopoles condensed. (More than $N-n$ condensed
monopoles correspond to those classical vacua in \higgsi\ where some
$N_i=0$, and thus there are fewer than $n$ photons left massless.)
The vacuum obtained from integrating out $u_p$ as in \WSWeom, 
will give some values of the $\ev{u_p}$ which are determined in terms
of the $g_p$.

Solving for the supersymmetric vacua as in \WSWeom, is equivalent to
minimizing $W_{tree}=\sum _{p=1}^{n+1}g_p u_p$, subject to the constraint
that $\ev{u_p}$ lie on the the codimension $N-n$ subspace of the
Coulomb branch where at least $N-n$ mutually local monopoles or dyons
are massless.  This is just a matter of replacing the monopoles with
$N-n$ Lagrange multipliers, imposing that the ${u_r}$ lie in the
subspace with $N-n$ massless monopoles; i.e. we integrate out the
$u_p$ with $W=W_{tree}+\sum _{k=1}^{N-n}L_kM_k(u)$, with $M_k(u)$ the
monopole masses on the Coulomb branch and $L_k$ Lagrange multipliers,
and the $\ev{L_k}=\ev{q_k\widetilde q_k}$.  The resulting $\ev{u_p}$
will be some fixed value, depending on the $g_r$ and $\Lambda$, giving
finally $W_{exact}(g_r, \Lambda) =\sum _{r}g_r\ev{u_r}$.

Recall that the curve of the $U(N)$ theory is
\eqn\SUNc{y^2=P(x;u_r)^2-4\Lambda ^{2N}, \qquad P(x,u_r)\equiv 
\det(x-\Phi)=\sum_{k=0}^Nx^{N-k}s_k,}
with the $s_k$ related to the $u_r$ by
\eqn\newton{ks_k+\sum _{r=1}^kru_rs_{k-r}=0,}
and $s_0\equiv 1$ and $u_0\equiv 0$; thus $s_1=-u_1$, $s_2= \half u
_1^2-u_2$, etc. (for $SU(N)$ we impose $u_1=0$).  The condition for
having $N-n$ mutually local massless magnetic monopoles is that
\eqn\mmonc{P_N(x;\ev{u_p})^2-4\Lambda ^{2N}=(H_{N-n}(x))^2F_{2n}(x),}
where $H_{N-n}$ is a polynomial in $x$ of degree $N-n$ and $F_{2n}$ is
a polynomial in $x$ of degree $2n$.  The LHS of \mmonc\ has $2N$
roots, and the RHS says that $N-n$ pairs of roots should be tuned to
coincide; thus \mmonc\ is satisfied on codimension $N-n$ subspaces 
of the Coulomb branch. We
need to integrate out the $u_p$, with $W_{tree}=\sum _{p=1}^{n+1} g_p u_p$,
subject to the constraint that $\ev{u_p}$ satisfy \mmonc. 

Of the $n$ massless photons, the one corresponding to the
trace of $U(N)$, does not couple to the rest of the theory
and so its coupling constant is the same as the one we started with.
The other $n-1$ photons which are left massless in \higgsi\ have gauge
couplings which are given by the period matrix of the reduced curve
\eqn\redc{y^2=F_{2n}(x; \ev{u_r})=F_{2n}(x; g_p , \Lambda ),}
with $F_{2n}(x;\ev{u_p})$ the same function appearing in \mmonc\ and
$\ev{u_r}$ the point on the solution space of \mmonc\ which minimizes
$W_{tree}$.  The curve \redc\ thus gives the exact gauge couplings of
$U(1)^{n-1}$ which remain massless in \higgsi\ as functions of 
$g_p$ and $\Lambda$.  

The dual Calabi-Yau geometry which we proposed in section 4,
$$ W'(x)^2+f_{n-1}(x)+y^2+z^2+v^2=0,$$
is already similar to the SW geometry \redc, giving the coupling
constants of the massless $U(1)$'s.  To show that the $\tau _{ij}$
obtained from \redc\ agrees with that obtained from \tauis, we
need to show that the $F_{2n}(x)$ of \mmonc\ and \redc\ is given by
\eqn\FWrel{g_{n+1}^2F_{2n}(x)=W'(x)^2+f_{n-1}(x),}  
with the factor of $g_{n+1}^{2}$ because the highest order term in
$F_{2n}(x)$ is $x^{2n}$, whereas that of $W'(x)$ is $g_{n+1}x^n$.  We will
indeed verify that the structure of $F_{2n}$ predicted from
 \FWrel\ is correct, i.e. it is a deformation of a degree $n-1$ polynomial
in $x$ added to $W'^2$.  However more needs to be done to show
that the dual geometry and gauge theory predict the same
coupling constants for the $U(1)$'s.  Namely, we have to show
that the coefficients of the $f_{n-1}$ predicted
{}from dual geometry and that of the gauge theory have identical
dependence on $N_i$ and the parameters of the superpotential.
This is indeed a highly non-trivial statement, which we will later
verify for cubic superpotential in section 8.

As a first hint about why \FWrel\ holds, consider the classical 
limit, $\Lambda \rightarrow 0$, where $P_N(x)=\det(x-\Phi)\rightarrow 
\prod _{i=1}^n (x-a_i)^{N_i}$, with $a_i$ the roots of $W'(x)=g_{n+1}
\prod _{i=1}^n (x-a_i)$.  In this limit $P_N^2-4\Lambda ^{2N}\rightarrow
H_{N-n}^2F_{2n}$, as in \mmonc, with $H_{N-n}(x)=\prod
_{i=1}^n(x-a_i)^{N_i-1}$ and $F_{2n}=\prod
_{i=1}^n(x-a_i)^2=g_{n+1}^{-2}W'(x)^2$.  The motivation for this splitting
is applying the intuition of \DS\ to each $SU(N_i)$ factor: each $P_{N_i}^2-1$
splits to $(x-a_i)^2$ times a degree $N_i-1$ polynomial.  We thus find 
that \FWrel\ holds in the $\Lambda\rightarrow 0$ limit, and see that the 
$f_{n-1}(x)$ appearing in \FWrel\ satisfies $f_{n-1}(x)\rightarrow 0$
for $\Lambda \rightarrow 0$.  

To prove \FWrel\ exactly, and also get some insight into how the $\ev{u_r}$
are determined, we note that 
we can minimize our $W_{tree}$ \wtreei, subject to the constraint that 
the $\ev{u_r}$ satisfy \mmonc, by introducing several Lagrange multipliers:
\eqn\wLM{W=\sum _{r=1}^n g_ru_r+\sum _{i=1}^{l}
[L_i(P_N(x;u_r)\big| _{x=p_i}-2\epsilon _i\Lambda ^N)+Q_i{\partial
\over \partial x} P_N(x;u_r)\big| _{x=p_i}],} with $\epsilon _i = \pm
1$.  We're generally allowing $l$ mutually local massless monopoles,
and will see that $l\geq N-n$. The $L_i$, $Q_i$, and $p_i$ are all
treated as Lagrange multipliers; so we should independently take
derivatives of
\wLM\ with respect to all $u_r$, $L_i$, $Q_i$, and $p_i$, and set all
these derivatives to zero.  The $p_i$ will be the roots of
$H_{l}(x)$ in \mmonc, and the $L_i$ and $Q_i$ constraints
implement the LHS of \mmonc\ having double zeros at these $l$ points
$p_i$.  

The variation of \wLM\ with respect to $p_i$ gives
\eqn\pivar{Q_i{\partial ^2 P_N\over \partial x ^2}\bigg| _{x=p_i}=0,}
where we used the $Q_i$ constraint to eliminate the term involving $L_i$.
For generic $g_r$, the RHS of \mmonc\ has some double roots, but no
triple or higher roots; therefore \pivar\ implies that $\ev{Q_i}=0$.
The situation where the RHS of \mmonc\ does have triple or higher order
roots is where the unperturbed $\N =2$ theory has an interacting
$\N =2$ superconformal field theory, as in \ArgDoug.  Our $\N =1$
theory with $W_{tree}$ does put the vacuum at such points for some
special choices of the $g_r$, but we'll consider the generic
situation for the moment.  Since the $\ev{Q_i}=0$, the variation of
\wLM\ with respect to all $u_r$ gives
\eqn\wLMur{g_r+\sum _{i=1}^{l}\sum _{j=0}^NL_ip_i^{N-j}{\partial s_j\over 
\partial u_r}=0,}
with the understanding that the $g_r=0$ for $r>n+1$.  Using \newton,
\wLMur\ becomes
\eqn\wLMurr{g_r=\sum _{i=1}^{l}\sum _{j=0}^NL_ip_i^{N-j}s_{j-r}.}
We should also impose the $L_i$ and $Q_i$ constraints in \wLM.  These
equations and \wLMurr\ fix the $\ev{u_r}$, $\ev{L_i}$, $\ev{p_i}$, and
$\ev{Q_i}$ as functions of the $g_r$ and $\Lambda$.  The $\ev{L_i}$
are proportional to the expectation values $\ev{q_i\widetilde q_i}$ of
the $l\geq N-n$ condensed, mutually local, monopoles.

Following a similar argument in \JBYO, we multiply \wLMurr\ by $x^{r-1}$
and sum: 
\eqn\Wderi{\eqalign{W_{cl}'(x)&=\sum _{r=1}^Ng_rx^{r-1}
\cr & =\sum _{r=1}^N\sum _{i=1}^{l}\sum_{j=0}^Nx^{r-1}p_i^{N-j}s_{j-r}L_i
\cr &=\sum _{r=-\infty }^N\sum _{i=1}^l\sum_{j=0}^Nx^{r-1}p_i^{N-j}
s_{j-r}L_i-2L\Lambda ^{N}x^{-1}+
{\cal O}(x^{-2})\cr
&=\sum _{i=1}^l\sum_{j=-\infty }^N P_N(x;\ev{u})x^{j-N-1}p_i^{N-j}
L_i-2L\Lambda ^N x^{-1}+{\cal O}(x^{-2})\cr
&=\sum _{i=1}^l{P_N(x;\ev{u})\over x-p_i}L_i-2L\Lambda ^Nx^{-1}
+{\cal O}(x^{-2}).}}
We define $L\equiv \sum _{i=1}^l L_i\epsilon _i$.
Defining, as in \JBYO, the order $l-1$ polynomial $B_{l-1}(x)$ by
\eqn\Bdefn{\sum_{i=1}^l{L_i\over x-p_i}={B_{l-1}(x)\over H_l(x)},}
with $H_l(x)$ the polynomial appearing in \mmonc,
we thus have
\eqn\wderii{W_{cl}'(x)+2L\Lambda ^Nx^{-1}=B_{l-1}(x)
\sqrt{F_{2N-2l}(x)+{4\Lambda ^{2N}\over H_l(x)^2}}+{\cal O}(x^{-2}).}
Since the highest order term in $W'_{cl}$ is $g_{n+1}x^n$, 
we see that $B_{l-1}(x)$ should
actually be order $n-N+l$.  This shows that $l\geq N-n$ and, in
particular, for $l=N-n$, $B_{N-n-1}=g_{n+1}$ is a constant.
Squaring \wderii\ gives
\eqn\squarei{g_{n+1}^2F_{2n}=
W_{cl}'^2+4g_{n+1}L\Lambda ^Nx^{n-1}+{\cal O}(x^{n-2}).}
We have thus derived \FWrel, $g_{n+1}^2F_{2n}=W'^2+f_{n-1}(x)$, and
found that $f_{n-1}(x)=4g_{n+1}L\Lambda ^Nx^{n-1}+{\cal O}(x^{n-2})$.

This shows that the exact $\tau _{ij}(g_r, \Lambda)$ of the $U(1)^n$
photons left massless found using the reduced $\N =2$ curve \redc,
evaluated in the supersymmetric vacua, is consistent with that of \tauis,
found in section 4 via our large $N$ duality.  However as noted before
to show they are exactly the same we have to match the coefficients
of $f_{n-1}(x)$, which depends in a highly non-trivial way on $N_i$
and the coupling constants of the superpotential.
 The above method also,
in principle, gives the $\ev{u_r}$, and thus $W_{eff}(g_r)$, which can
be compared with the duality result $W_{exact}(g_r, S_i)$
\suppo\ (upon integrating out the $S_i$).
The duality results  
\tauis\ and \suppo\ give the answers, and in particular the 
$N_i$ dependence, in a much simpler and more elegant fashion.

It is interesting to ask if the duality results of section 4 could be
recovered more directly by a field theory analysis which includes the
$n$ glueball chiral superfields $S_i$ of the unbroken gauge group
$\prod _{i=1}^nU(N_i)$.  In the original $SU(N)$ theory, we can construct
$N$ generalized glueball objects $\sim \Tr \Phi ^{i}W_\alpha W^\alpha$,
$i=0\dots N-1$.  The $N-n$ monopole condensates or Lagrange multiplier
expectation values in the above analysis is (indirectly) related to $N-n$
of these generalized glueballs.  The $n$ remaining ones should be those
of the unbroken low-energy $\prod _{i=1}^nU(N_i)$.  It is not known how
to exactly include these from a direct field theory analysis.

For any $W_{tree}$, there are vacua where classically $U(N)$ or
$SU(N)$ is unbroken and, in the quantum theory, $N-1$ mutually local
monopoles condense.  These are the only vacua for $W_{tree}=mu_2$, but
also exist for any $n\geq 1$.  The condition for having the $N-1$
mutually local massless monopoles is \DS
\eqn\mmoncc{P(x;\ev{u_r})^2-4\Lambda ^{2N}=H_{N-1}(x)^2 F_2(x),}
which is satisfied via Chebyshev polynomials:
\eqn\chebi{P_N(x,\ev{u_r})=\Lambda ^NT_N({x\over \Lambda}); 
\qquad T_N(x\equiv t+t^{-1})=t^N+t^{-N}.}
With the normalization of \chebi, $T_N(x)=x^N-Nx^{N-2}+\dots$, the
first Chebyshev polynomials.  The roots of $P_N=\det (x-\Phi)$, as
given by \chebi, are $\phi _j=2\Lambda
\cos ((2j+1)\pi /2N)$, $j=0\dots N-1$; this gives \wexacti.

More generally, we can use Chebyshev polynomials to construct new
solutions of the massless monopoles constraint \mmonc.  
Given a solution $P_N(x)$ of \mmonc\ which is appropriate for the
$SU(N)$ theory where the vacuum is broken to
\eqn\higgsBb{SU(N)\rightarrow \otimes _{i=1}^nSU(N_i)\otimes U(1)^{N-1}
\qquad
\hbox{with}\quad\sum _i N_i=N,}
we can immediately construct the solution $P_{KN}(x)$ which is
appropriate for a $SU(KN)$ theory, with the same $W_{tree}$
\wtreei, in the vacuum where the gauge group is broken as
\eqn\higgsbb{SU(KN)\rightarrow 
\otimes _{i=1}^nSU(KN_i)\otimes U(1)^{N-n}\qquad
\hbox{with}\quad\sum _i N_i=N.}  The solution $P_{KN}(x)$ of \mmonc\ for the
theory \higgsbb\ is given by the Chebyshev polynomial of the $K=1$
solution $P_N(x)$:
\eqn\chebg{P_{KN}(x)=\widetilde \Lambda ^{NK}T_K\left( {P_N(x)\over
\Lambda ^N}\right) ,}  
with $\widetilde \Lambda$ and $\Lambda$ the scales of $SU(KN)$ and
$SU(N)$, respectively. 
To see that this satisfies the condition of \mmonc\ note
\eqn\satis{\eqalign{P_{KN}(x)^2-4\widetilde \Lambda ^{2KN}&=\widetilde 
\Lambda ^{2NK}(T_K\left( {P_N\over \Lambda ^N}
\right) ^2-4)=\widetilde \Lambda ^{2KN}[U_{K-1} \left( {P_N\over \Lambda
^N}\right) ]^2({P_N^2\over \Lambda ^{2N}}-4)\cr
&=\widetilde \Lambda ^{2KN}
\Lambda ^{-2N}[U_{K-1}\left( {P_N\over \Lambda ^N}\right)
H_{N-n}(x)]^2 F_{2n}(x)
\equiv [H_{KN-n}(x)]^2F_{2n}(x).}}
We denote the second
Chebyshev functions $U_{K-1}(x\equiv t+t^{-1})\equiv
(t^K-t^{-K})/(t-t^{-1})=x^{K-1}+\dots$,
and the second line uses the fact that $P_N$ is a  
solution of \mmonc.  Thus 
$P_{NK}(x)$ given by \chebg\ indeed satisfies the
condition \mmonc\ appropriate for \higgsbb. Furthermore, 
the $U(1)^{N-n}$ in \higgsbb\ has gauge couplings given by the curve
$y^2=F_{2n}(x)$, which is the same as that of the $K=1$ theory. This
fits with the dual geometry prediction of section 4, as will be discussed
in the next section.

Expanding out \chebg\ relates the expectation values $\ev{\widetilde u _p}$
of the $SU(KN)$ theory to the $\ev{u_p}$ of the $SU(N)$ theory.  The
relation is especially simple for the lower Casimirs:
\eqn\uvK{\widetilde u_2=Ku_2, \qquad \widetilde u_3=Ku_3, }
with some more complicated relations for the general higher Casimirs.

By the above construction, it suffices to consider \higgsi\ where the
$N_i$ have no common integer divisor.  The simple $K$ dependence fits
with the duality results of section 4.

\subsec{Other possible connections}

The quantum $\N =2$ theory is related to an integrable hierarchy,
which is known to have integrable ``Whitham hierarchy deformations;''
see e.g. \whitham.  Our superpotential $W_{tree}$ is naturally
regarded as a Whitham deformation of the $\N =2$ theory, where the
Whitham ``times'' are the $g_r$ in \wtreei\ which, from the $\N =2$
perspective, are spurions breaking $\N =2$ to $\N =1$.  The exact
solution can still be obtained as a $\Theta $ function of the Whitham
hierarchy, see e.g. the last reference of \whitham.  It would be
interesting to see how this $\Theta$ function is related to the $S_i$
and $\Pi _i$ periods of section 4.

The $\N =1$ $U(N)$ field theories with adjoint $\Phi$, $N_f$
fundamental flavors, and general superpotential $W_{tree}(\Phi)$
\wtreei\ can also be constructed via $N$ IIA D4 branes suspended between
a NS brane and $n$ NS' branes.  The construction was discussed in
detail in \JBYO\ and references cited therein.  Four of the five
directions transverse to the D4s in IIA are conventionally written as
having complex coordinates $w$ and $v$.  The NS' branes are given by
some $(v,w)$ curve, which classically is $w=W_{tree}'(v)$, giving the
$n$ NS' branes at the minima of $W_{tree}$.  Going to M-theory, the
brane configuration becomes a smooth M5 brane configuration, as in
\wittens.  Our geometric flop transition duality is roughly reminiscent of
exchanging the roles of $v$ and $w$; it was already speculated \JBYO\ that
this exchange could be related to the field theory duality of \kutetal.
Perhaps this can be made more precise.

\newsec{The case with the cubic superpotential in more detail}

Consider in more detail the case $n=2$, with $W_{cl}=gu_3+mu_2+\lambda
u_1$.  Then $W'=g(\phi - a_1)(\phi - a_2)$, with
\eqn\ai{a_1={m\over 2g}+\sqrt{\left({m\over 2g}\right)^2-
{\lambda \over g}}, \quad a_2=
{m\over 2g}-\sqrt{\left({m\over 2g}\right)^2-{\lambda \over g}}.}
For $SU(N)\rightarrow SU(N_1)\times SU(N_2)\times U(1)$, as opposed to
$U(N)\rightarrow U(N_1)\times U(N_2)$, $\lambda$ should be treated as a
Lagrange multiplier, enforcing $u_1=0$.  In that case,
\eqn\aii{a_1=\left({m\over g}\right){N_2\over (N_1-N_2)}, 
\quad a_2=-\left({m\over g}\right)
{N_1\over (N_1-N_2)}.}
The classical low-energy superpotential is
\eqn\Wclii{W_{cl}={m^3\over g^2}\cdot {N_1N_2(N_1+N_2)\over 6(N_1-N_2)^2}} and 
\eqn\matchii{\Lambda _1^{3N_1}=g^{N_1} \Delta ^{N_1-2N_2}\Lambda ^{2N}\qquad 
\Lambda _2^{3N_2}=g^{N_2}\Delta ^{N_2-2N_1}\Lambda ^{2N},}
with $m_W=a_1-a_2=(m/g)(N/(N_1-N_2))\equiv \Delta $ and $m_\phi =g\Delta$.
Naive ``integrating in'' then gives $W_{eff}=W_{cl}+W_{np}$ with 
\eqn\Wiiinii{\eqalign{W_{np}&=\sum_{i=1}^2
S_i\left( \log({\Lambda _i^{3N_i}\over S_i^{N_i}})+N_i\right)\cr
&=N_1\left[S_1\log({g \Lambda ^2\Delta \over S_1})+S_1 +S_2
\log({\Lambda ^2\over \Delta ^2})\right]
+N_2\left[S_2\log({g \Lambda ^2\Delta \over S_2})+S_2+ S_1
\log({\Lambda ^2\over \Delta ^2})\right].}}

The exact answer for the {\it value}
of the superpotential at the minima of $W$ can be
 obtained via deforming the ${\cal N}=2$ curve, is
given by \WSWl, with the $\ev{u_r}$ given by solving \mmonc\ for $n=2$:
\eqn\mmoncii{P_N^2-4\Lambda ^{2N}=H_{N-2}^2F_4.}  Again, this does
not include the glueball fields.

As discussed in the previous section, a solution of \mmoncii\ appropriate
for $SU(N)\rightarrow SU(N_1)\times SU(N_2)\times U(1)$ can be used to
immediately construct a solution of \mmoncii\ appropriate for 
$SU(KN)\rightarrow SU(KN_1)\times SU(KN_2)\times U(1)$.  Using \uvK, the
low energy effective superpotential for the $SU(KN)$ theory is
\eqn\wkfact{W_{eff}[SU(KN)]=m\ev{\widetilde u_2}+g\ev{\widetilde u_3}=Km\ev{u_2}+ Kg\ev{u_3}=KW_{eff}[SU(N)],}
simply a factor of $K$ times that of the $SU(N)$ theory.  The $\ev{c_l}$ which
minimizes $W_{eff}$, giving the vacuum on the solution space of \mmoncii, is
thus $K$ independent, so $K$ really does just factor out as an overall
multiplicative factor in the superpotential.

\subsec{Examples:}

{\bf $U(3N)\rightarrow U(2N)\times U(N)$}

\vglue 1cm

As a simple example of the procedure outlined in the last section,
consider the case of $U(3N)$ in the vacuum where the unbroken group is
$U(2N)\times U(N)$.  As discussed above it suffices to consider
the case $N=1$.  The superpotential of \wLM\ is
\eqn\Wexamp{W=\lambda u_1+mu_2+gu_3+L(p^3+s_1p^2+s_2p+s_3\pm 2\Lambda ^3)+
Q(3p^2+2s_1p+s_2).}
The $p$ equation of motion (along with $Q$'s) gives $\ev{Q}=0$ and 
\wLMurr\ then gives $\lambda =L(p^2+ps_1+s_2)$, $m=L(p+s_1)$, $g=L$.
Thus $\ev{s_1}=g^{-1}m-p$, $\ev{s_2} =g^{-1}(\lambda -mp)$, and
$\ev{s_3}=\mp 2\Lambda ^3-pg^{-1}\lambda$. $\ev{p}$ is fixed by the
$Q$ constraint to be either $a_1$ or $a_2$ of \ai, so 
$W_{cl}'(x)=g(x-p)(x+p+g^{-1}m)$.
We then have $\ev{P_3(x)}=g^{-1}(x-p)W_{cl}'(x)\mp 2\Lambda ^3$, and
thus $P_3^2-4\Lambda ^6=(x-p)^2F_4(x)$, with $g^2F_4(x)=W'(x)^2\mp
4g\Lambda ^3(gx+gp+m)$, which matches with \squarei.
For $SU(3)$, we treat $\lambda$ also as
a Lagrange multiplier, enforcing $\ev{s_1}=-\ev{u_1}=0$,
i.e. $\ev{p}=m/g$.  The $Q$ constraint then gives $\ev{\lambda }=-2m^2/g$,
so $\ev{u_2}=3(m/g)^2$ and $\ev{u_3}=-2(m/g)^3 \pm 2\Lambda ^3$.
Plugging these back into $W$ gives $W_{low}=(m^3/g^2)\pm 2g\Lambda ^3$.

Equivalently, we could simply solve the $L$ and $Q$ constraints at
the outset by taking $P_3=(x-a)^2(x-b)\mp 2\Lambda ^3$, giving
$\ev{u_1}=2a+b$, $\ev{u_2}=2a^2+b^2$, $\ev{u_3}=2a^3+b^3\pm 2\Lambda
^3$ and thus $W_{low} = 2 W(a)+W(b)\pm 2g\Lambda^3$.  Minimizing with
respect to $a$ and $b$ gives $\ev{a}=a_1$, $\ev{b}=a_2$ and
$W_{low}=W_{cl} \pm 2g\Lambda ^3$ with $W_{cl}=2W(a_1)+W(a_2)$.  In
order to get the $SU(3)\rightarrow SU(2)\times U(1)$ answer we impose
$\partial W_{low}/\partial \lambda=0$, which implies $a_1={m\over g}$,
$a_2 =-2{m\over g}$.

We thus find for $SU(3)$ $W_{low}=(m^3/g^2)\pm 2g\Lambda^3$ and the
remaining massless photon has gauge coupling $\tau (g\Lambda /m)$
which is given exactly by the curve $y^2=g^2F_4(x)= W'^2\mp 4g\Lambda
^3 (gx+2m)$, with $W'(x)=g(x-{m\over g})(x+2{m\over g})$.  This curve
degenerates at $(m/g)^3=\pm \Lambda ^3$, i.e. $\ev{u_3}=0$, which is
where an additional magnetic monopole becomes massless in the ${\cal
N}=2$ theory.  The $SU(2)$ glueball has $\ev{S}=\pm g\Lambda ^3$.

\vglue 1cm

{\bf Splittings of $SU(5)$}

\vglue 1cm

The computation of the one parameter family of $\N =2$ curves for
the different splittings of $SU(5)$, namely, $SU(3)\times SU(2)\times U(1)$
and $SU(4)\times U(1)$ can be done explicitly. This will provide the 
highly non-trivial exact
answer for the low energy effective superpotential that will be used to
check the answer from the geometry in section 8.4. As discussed before
this answer also provides the solution for $SU(5K)\rightarrow SU(3K)\times
SU(2K)\times U(1)$ and  $SU(5K)\rightarrow SU(4K)\times
SU(K)\times U(1)$ for any integer $K$.

We need to solve \mmoncii\ for $N=5$, i.e. to find $P_5(x)$ such that
\eqn\SWexam{P^2_5(x)-4\Lambda^{10} = F_4(x)H^2_3(x)}
Clearly, $P_5(x)$ has five parameters, given by the positions of the roots
since the coefficient of $x^5$ can be normalized to one.  However, three
of them have to be used to produce the three double roots and one in order
to impose the quantum tracelessness condition, i.e., to set to zero the
$x^4$ coefficient. This leaves us with a one parameter family of curves.

Let us set $\Lambda^5={1\over 2}$ and $H_3(x)=(x-a)(x-b)x$. The
LHS of \SWexam\ can be factored as $(P_5-1)(P_5+1)$ where it is clear
that the two factors should contain no common roots. Therefore we can
freely set,
\eqn\poly{P_5(x) = (x-a)^2(x-b)^2(x-c)\mp 1}
Now we want to make sure that $P_5\mp 1$ will have a double root at
$x=0$. This condition can be easily implemented by,
$$
P_5(0)=\pm 1   \qquad  {d P_5\over dx}(0)=0
$$
In terms of $a,b$ and $c$, these conditions read as follows,
\eqn\const{a^2b^2c=\pm 2  \qquad  a b (2c (a+b)+a b)=0}
Finally, we can impose the tracelessness condition by shifting
$x\rightarrow x-{1\over 5}(2(a+b)+c)$.  We can now read off the gauge theory
Casimir expectation values (using $\ev{Tr \Phi }=0$),
$$
P_5(x) =\ev{det(x-\Phi)}=
 x^5 - {1\over 2}\ev{Tr\Phi^2}x^3-{1\over 3}\ev{Tr\Phi^3}x^2+\ldots
$$

Since, our solution is symmetric in $a$ and $b$ it is more natural to write
it in term of the symmetric polynomials $s=a+b$ and $k =a b$. The
constraints \const\ now read $k^2 c=\pm 2$ and $k(2cs+k)=0$.  Assuming that
$k\neq 0$ we can solve for $k$ as $k =-2 cs$. Then we are left with only one
constraint, namely, $2 s^2 c^3 =\pm 1$.   

The Casimirs are now given by,
$$
u_2 = {1\over5}(2c^2+18 c s+3s^2 ) \qquad  u_3 =-{2\over 25}(2 c^3-23 c^2
s+9 c s^2+s^3)
$$
and the superpotential is now a function of $c$ or $s$ depending on how we
use the constraint. Let us introduce the constraint through a Lagrange
multiplier $\beta$ and write the superpotential as,
$$
W_{eff}(c,s,\beta) = g u_3(c,s)+m u_2(c,s)+\beta (\pm \Lambda^5-s^2 c^3)
$$
where we have introduced $\Lambda$ back for later convenience.

Now we need to solve ${\partial W_{eff}\over \partial c}=0$ and
${\partial W_{eff}\over
\partial s}=0$ and then impose the constraint. Computing these two equations
and using one of them to eliminate $\beta$ from the other we get the
following simple equation,
\eqn\impor{3c+s = {5m\over g}}
subject to the constraint  $s^2 c^3=\pm\Lambda^5$. There is yet a better
way to write the constraint, namely, $s^4 c^6 = \Lambda^{10}$. This will
make very simple the identification of the different vacua.

Now we can see how the different splittings will come out. The
classical limit corresponds to setting $\Lambda \rightarrow 0$ and the
constraint can be solved in two ways, namely, $s=0$ or $c=0$.  The
former leads to $c={5m\over 3g}$ using
\impor\ while the latter leads to $s= {5m\over g}$. Plugging this in the
superpotential we reproduce in the former case the classical answer for
$SU(4)\times U(1)$ and in the latter we get that of $SU(3)\times
SU(2)\times U(1)$. 

$SU(4)\times U(1)$

In order to get $W_{low}$ we need to solve for $c$ using
\impor\ and $s^4={\Lambda^{10} \over c^6}$. Clearly, we have 4 solutions to
the constraint giving $s=s(c)$. These are the $N_1 N_2=4$ vacua. The
equation we need to solve is then
$$
c= {5m\over 3g}-{s(c)\over 3}
$$
this can be solved recursively using $t\equiv ({3g\Lambda\over 5m})^{5/2}$ as
expansion parameter. Once this is done, $s$ can also be found and plugging
them back in the superpotential we get,
$$
\eqalign{W_{low}={125\over 27}{m^3\over g^2}\left( \right.& {2\over 25}+
4t-{1\over 3}t^2-{7\over
54}t^3-{5\over 54}t^4 -{221\over 2592}t^5-{22\over 243}t^6+\cr 
& \left. -{2185\over 20736}t^7-{286\over 2187}t^8-{9147325\over 53747712}t^9
+\ldots \right)}
$$
The above exact answer for the value of the superpotential
at the critical point differs from the naive integrating in analysis
\Wlows, which would terminate at order $t^2$.  The coefficients
of the classical $t^0$ term and $t$ term agree with the exact answer
above, but the coefficient of $t^2$ term differs from the exact
answer.

$SU(3)\times SU(2)\times U(1)$

In this case we need to solve for $s$ using $c^6={\Lambda^{10} \over
s^4}$. Here, we have 6 solutions giving the $N_1 N_2=6$ choices of
vacua. The equation in this case becomes,
$$
s = {5m\over g}-3c(s)
$$
solving as before but using as expansion parameter $t\equiv ({g\Lambda\over
5m})^{5/3}$ we get for the superpotential the following expression,
$$
\eqalign{W_{low}= {250\over 2}{m^3\over g^2}\left( \right.&\left. {1\over
25}+3t^2+2t^3+6t^4+26t^5+135t^6+782t^7+{14630\over 3}t^8 + \right.\cr
& \left. +32076 t^9 +\ldots \right)}
$$
Again this differs from the result of the naive low energy analysis
\Wlows\ which would terminate at order $t^3$; up to that order
the naive answer agrees with the above exact answer.

\vglue 1cm

{\bf Splitting  $U(5)\rightarrow U(3)\times U(2)$}

\vglue 1cm

It is also possible to find the curve for $U(5)$ and from it to
compute the $SU(5)$ answer by imposing the tracelessness
constraint. However, the computation for $U(5)$ is more cumbersome than the
$SU(5)$ counterpart. In this part of the section we will simply show the
answer for the low energy effective superpotential and the computation can
be found in Appendix A.

Since we now do not impose the tracelessness condition, $\lambda$ is a 
free parameter, rather than a Lagrange multiplier. $\lambda /g$, $m /g$ and
$\Lambda$ combine into a single expansion parameter 
$$
T^3=\left({\Lambda\over \Delta}\right)^5,
$$  
with $\Delta =a_1-a_2=\sqrt{({m\over g})^2-4{\lambda\over g}}$.
The low energy superpotential is then given by,
$$
W_{low} = 3 W(a_1)+2 W(a_2)+ g \Delta^3 \left(3T^2+2T^3+4T^4+10T^5+
\ldots \right).$$
In the dual geometric picture we will see that $U(5)$ is the natural
answer obtained, and then one has to impose the constraint to get the
$SU(5)$ superpotential.

\newsec{The analysis of the dual geometry}

The dual geometry proposal gives rise to the superpotential of section 4.1:
\eqn\superpow{-\Tpi W_{eff}=\sum_{i=1}^n N_i \Pi_i+\alpha (\sum_{i=1}^{n} 
S_i),}
where $\Pi_i$'s are the periods of the dual cycles and the
$S_i$'s are the sizes of the $S^3$'s as defined in \fperiod\ and \dualperiod. 

Using \fperiod\ and \dualperiod, it is seen that under $\Lco \rightarrow 
e^{2\pi
i}\Lco$ the $\Pi_i$ period will change by,
\eqn\sign{\Delta \Pi_i = - 2(\sum_{j=1}^{n}\pm S_j).}
The factor of two comes from the fact that we are dealing with
two copies of the x-plane connected by the $n$ branch cuts. (See Figure 2) 
Let us choose the orientation of the fundamental periods to be clockwise,
therefore, it is easy to see that we always get the upper sign in \sign\
for all $i$ and $j$.  We thus see that, in general, 
$\Pi_i$ must depend on the cutoff $\Lco$ as
\eqn\unex{\Pi_i = -{2\over 2\pi i}(\sum_{j=1}^n S_j)\log \Lco + \ldots ,}
with $\dots$ single valued under $\Lambda _0\rightarrow e^{2\pi
i}\Lambda _0$.

We now consider the full $\Lco$ dependence. Consider the region of
integration where $x$ is large compared to all $a_i$'s. Therefore we
can expand the effective one-form $\omega$ in $x$ around $x=\infty$
and it is easy to see that,
$$
\omega =\sqrt{W'(x)^2+f_{n-1}}dx = \left( W'(x)+ 
{1\over 2g_{n+1}}{b_{n-1}\over x}+{\cal O}({1\over x^2}) \right) dx
$$
where $b_{n-1}$ is the coefficient of $x^{n-1}$ in the deformation
polynomial $f_{n-1}(x)$ and
$W'(x)=g_{n+1}\prod_{j=1}^n(x-a_j)$. Integrating this we get,
\eqn\unextwo{\Pi_i = \ldots + W(\Lco )+{b_{n-1}\over 2g_{n+1}}\log{\Lco} 
+ {\cal O}({1\over \Lco })}  
where $\ldots$ are the $\Lco$ independent pieces.
This allows us to make the following identification using \unex\ and \unextwo.
$$
b_{n-1} = - 4 g_{n+1}\sum_{j=1}^n S_j.
$$
Comparing with \squarei, we see that we must have $\sum
_j\ev{S_j}=-L\Lambda ^N$, where both sides can be solved for in terms
of the $g_r$ and $\Lambda$.  As mentioned in section 4.1, $W(\Lco )$
is an irrelevant constant that can be ignored. However, we have to
deal with the logarithmic dependence because we want to take
$\Lco\rightarrow \infty$ at the end.  Notice that, had we included
deformations of degree higher than $n-1$, more singular divergences
would have appeared in \unextwo\ that do not have a counterpart in the
gauge theory side.  This shows again that, as in \squarei, the
deformation $f$ in $F\sim W'^2+f$ must have degree at most $n-1$.

Since every $\Pi_i$ has the same logarithmic divergence we can write the
contribution to the superpotential as follows,
$$
W_{eff}=\ldots + 2(\sum_{i=1}^n N_i)(\sum_{j=1}^nS_j)\log \Lco - 
2\pi i\alpha (\sum_{k=1}^nS_k)
$$
Now it is clear that the only way to obtain finite expressions is to
take $\alpha$ depending on $\Lambda _0$ such that 
\eqn\combine{N \log \Lambda = N \log \Lco - \pi i\alpha}
is finite.  Using $\sum_{j=1}^n N_j =N$, we can replace $\Lambda_0$ in
$W_{eff}$ by the physical scale $\Lambda$ of the $SU(N)$ theory.
 
Note that, for fixed $\Lambda$, the
superpotential for a splitting of the form $KN\rightarrow \sum_{i=1}^nKN_i$
has a trivial $K$ dependence: 
$$
-\Tpi W_{eff} = \sum_{i=1}^n KN_i \Pi_i = K (\sum_{i=1}^nN_i\Pi_i)
$$
if we replace $\Lco$ by $\Lambda$ in the $\Pi_i$'s by using the $\alpha$ term.
This matches with the results obtained from the gauge theory solution
\chebg\ using Chebyshev polynomials.

Some of the $S_i$ dependence of $\Pi_i$ can also be determined by
using monodromy arguments. Consider the semiclassical
regime, $\mid a_i^+-a_i^-\mid \ll \mid a_j-a_k\mid$  for all
$i,j,k$. Recall that
$W'(x)^2+f_{n-1}(x)=g^2_{n+1}\prod_{k=1}^n(x-a_k^+)(x-a_k^-)$. 
In this regime $S_i$ can be written as follows,
$$
S_i = \Tpi W''(a_i)\int^{a_i^+}_{a_i^-}\sqrt{(x-a_i)^2-\mu_{eff}}dx 
$$
where we have Taylor expanded $W'(x)^2 + f_{n-1}(x)$ around $x=a_i$ and
$$
\mu_{eff}\equiv -{1\over W''(a_i)^2}(f_{n-1}(a_i)+\ldots ). 
$$
Each $S_i$, in this limit, has been reduced to that
of the single conifold, which has
$$
S_i =  W''(a_i) \mu_{eff}
$$
up to a numerical coefficient. On the other hand, it is easy to see
that under $\mu_{eff}\rightarrow e^{2\pi i}\mu_{eff}$, $\Pi_i$ changes
by $\Delta\Pi_i=S_i$.  Therefore we conclude that,
$$
\Pi_i = \Tpi S_i \log \mu_{eff} \ldots 
=\Tpi S_i \log {S_i \over W''(a_i)} +\ldots
$$

Finally, we want to consider what happens to $\Pi_i$ when we move the
$j$-th 3-sphere all the way around the $i$-th 3-sphere. This corresponds to
changing $\Delta_{ij} = a_i-a_j$ to $e^{2\pi i}\Delta_{ij}$ leaving $a_i$
fixed. Under this operation we get $\Delta\Pi_i= 2 S_j$ (see
Figure 3). Therefore, 
$$
\Pi_i = \ldots + {2\over 2\pi i}\sum_{j\neq i}S_j \log\Delta_{ij}.
$$

\bigskip
\centerline{\epsfxsize=0.85\hsize\epsfbox{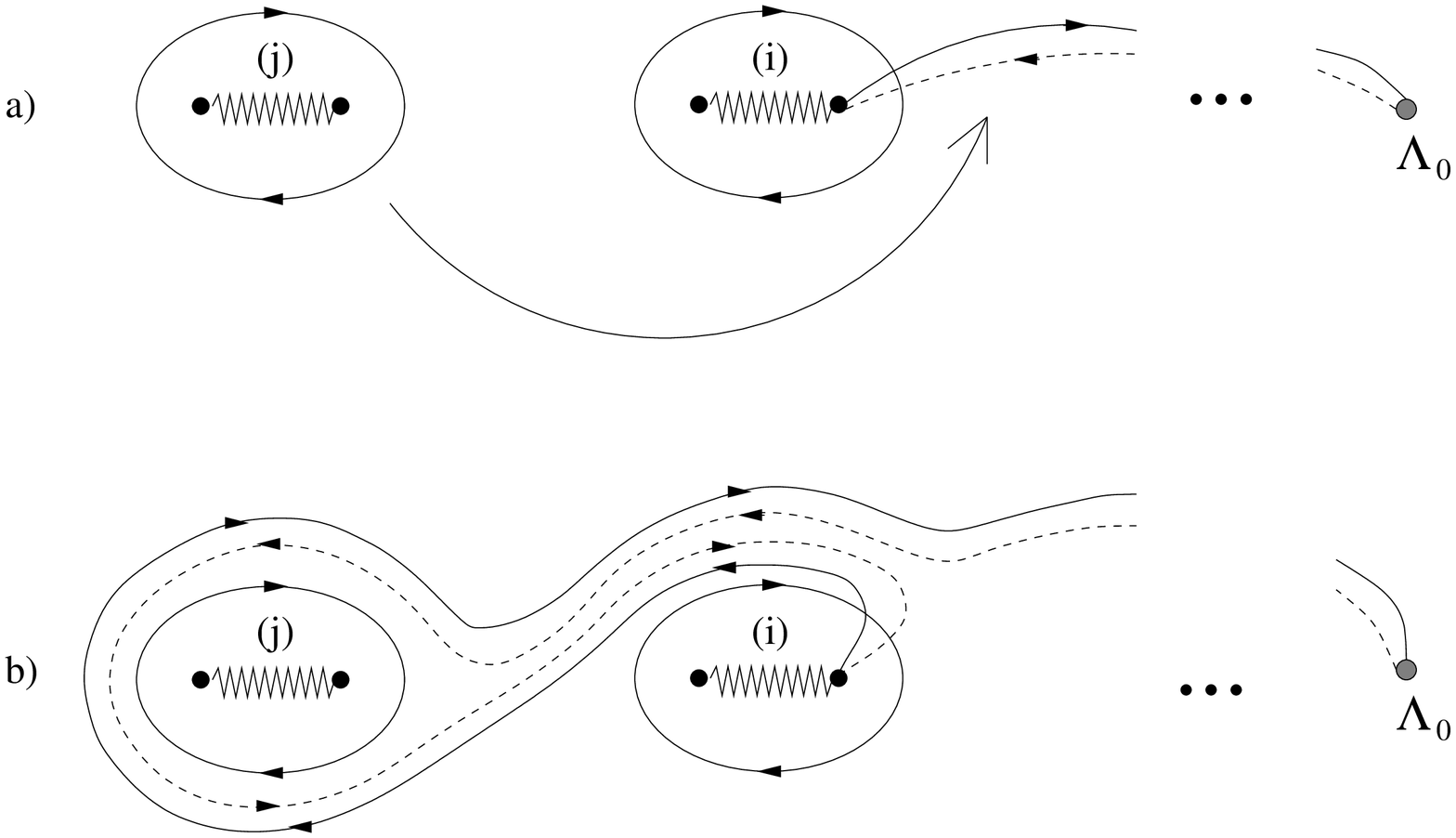}}
\noindent{\ninepoint\sl \baselineskip=8pt {\bf Figure 3}: {\sl a) Contours
of integration for $S_j$, $S_i$ and $\Pi_i$ before moving the $j$-th $S^3$
around the $i$-th $S^3$. b) The $\Pi_i$ contour goes around the $j$-th
sphere after the operation in a). }}
\bigskip

Now we can collect all these partial results in order to write,
$$
2\pi i\Pi_i = S_i \log {S_i \over W''(a_i)} + 2\sum_{j\neq i}S_j
\log\Delta_{ij}-2\sum_{k=1}^n S_k \log\Lco +\ldots.
$$
Plugging this back in \superpow\ and collecting all the $S_i$ pieces, we
get
$$
W_{eff} = \sum_{i=1}^n S_i \log\left( {W''(a_i)^{N_i}\prod_{j\neq
i}\Delta_{ij}^{-2N_j} \Lambda^{2N}\over S_i^{N_i}} \right) +\ldots, 
$$
with the $\dots$ single valued.

Comparing this to \Wiin\ and \lamin\ we see that we have re-derived
the approximate $W_{eff}$ obtained in section 5.1 as well as the naive
threshold matching relations. However, the above analysis can not rule out
further corrections to each $\Pi_i$ and hence to $W_{eff}$ in the form of a
power series in $S_i$'s. Indeed, as we will discuss in detail
for the case of the cubic superpotential, there is generally an infinite
power series in $S_i$'s which corrects the above expression.

\newsec{Cubic superpotential from geometry: An explicit computation}

In this section we consider the $n=2$ case, deforming the $\N=2$
theory by $W_{tree}=\lambda u_1+mu_2+gu_3$.  This was discussed in
detail from the gauge theory perspective in section 6.  We now focus
on the geometry side of the duality. In order to get the contribution
of the fluxes to the superpotential, we need to compute the periods of
the relevant cycles in the geometry.  For this $n=2$ case, \superpow\ 
gives
\eqn\effg{-\Tpi W_{eff} = N_1 \Pi_1 + N_2 \Pi_2 + \alpha (S_1 + S_2).} 
The fundamental periods are given as in \fperiod\ by,
\eqn\fund{S_1=\Tpi\int^{x_4}_{x_3}\omega  
\qquad  S_2=\Tpi\int^{x_2}_{x_1}\omega}
and the dual periods by
\eqn\dual{\Pi_1=\Tpi\int^{\Lco }_{x_3}\omega  \qquad  
\Pi_2=\Tpi\int^{x_1}_{-\Lco }\omega}
where we have denoted by $x_i$ the roots of the quartic polynomial 
$W'(x)^2+f_1(x)$
appearing in the definition of the effective one-form instead of
$a_i^+,a_i^-$ as in last section, in order to simplify the notation.

To compute the effective superpotential, we need to express the dual
periods $\Pi _1$ and $\Pi _2$ in terms of the fundamental periods $S_1$
and $S_2$.  Since, on the gauge theory
side, one does not have the exact answer for the superpotential
in terms of the glueball fields, we need to integrate out the $S_i$,
fixing them at their supersymmetric vacua $\ev{S_i}$.
This will give $W_{exact}(\lambda, m, g, \Lambda)$, which can be compared with
the gauge theory results.

Recall that $\lambda$ is a free parameter only for the $U(N)$ theory.
For $SU(N)$, which we will also compare, $\lambda$ is a Lagrange
multiplier imposing (quantum) tracelessness; this will fix 
$\lambda$ in terms of $m$, $g$ and $\Lambda$ and the $N_i$.

\subsec{Computation of the periods} 

As discussed in the general case in section 7, only by using monodromy
arguments it is possible to show the general form of the $S_i$ dependence of
the dual periods. In our case, this reads,
\eqn\dualone{\Pi_1 =\Tpi\left( W(\Lco )-W(a_1) + S_1\log{S_1\over
g\Delta} - S_1 + 2S_2\log\Dph -2(S_1+S_2)\log\Lco + P \right) }
where $P=P(S_1,S_2)$ is an infinite power series in $S_1$ and $S_2$,  $\Dph
=a_1-a_2$ and $W(x) = (1/3)g x^3+(1/2)mx^2+\lambda x$. Recall that
$W'(x)=g(x-a_1)(x-a_2)$ was introduced in section 6. Use has also been
made of $W''(a_1)=g\Delta$.

The explicit computation of $P(S_1,S_2)$ can be found in Appendix B up
to order $S_i^4$ where a method to compute higher order contributions is also
given. Here we will only show the result for $\Pi_1$ and $\Pi_2$ that will
be used later in this section.  
$$
\eqalign{2\pi i\;\Pi_1 =&W(\Lco)-W(a_1)+S_1 (\log{S_1\over g\Delta} - 1) 
+ 2S_2\log\Dph - 2
(S_1+S_2)\log \Lco + \cr
&+ g(\Dph)^3 \left[{1\over (g\Dph^3)^2}\left( 2S_1^2-10S_1 S_2+5 S_2^2\right)
+ {1\over (g\Dph^3)^3}\left( {32\over 3}S_1^3-91S_1^2S_2 +\right. \right.\cr
&\left. +118S_1S_2^2-{91\over
3}S_2^3\right) 
 + {1\over (g\Dph^3)^4}\left({280\over 3}S_1^4-{3484\over
3}S_1^3S_2+2636S_1^2S_2^2+\right. \cr
&\left.\left. -{5272\over 3}S_1S_2^3+{871\over 3}S_2^4 \right)+
{\cal O}\left({S^5 \over (g\Dph^3)^5}\right) \right]}
$$
and,
$$
\eqalign{2\pi i\;\Pi_2 =&W(-\Lco)-W(a_2)+S_2 (\log{S_2\over g\Delta} - 1) 
+ 2S_1\log\Dph - 2
(S_1+S_2)\log \Lco + \cr
&- g(\Dph)^3 \left[{1\over (g\Dph^3)^2}\left( 2S_2^2-10S_1 S_2+5 S_1^2\right)
- {1\over (g\Dph^3)^3}\left( {32\over 3}S_2^3-91S_2^2S_1 +\right. \right.\cr
&\left. +118S_2S_1^2-{91\over
3}S_1^3\right) 
 + {1\over (g\Dph^3)^4}\left({280\over 3}S_2^4-{3484\over
3}S_2^3S_1+2636S_2^2S_1^2+\right. \cr
&\left.\left. -{5272\over 3}S_2S_1^3+{871\over 3}S_1^4 \right)+
{\cal O}\left({S^5 \over (g\Dph^3)^5}\right) \right]} 
$$ 

\subsec{Low Energy Superpotential}

In order to compute the low energy superpotential we have to integrate out
$S_1$ and $S_2$ from the effective superpotential. In order to do this in
practice, it is convenient to define 
$$
x\equiv {S_1 \over g\Dph^3} \quad  y \equiv {S_2 \over (-g\Dph^3)} 
$$

In term of these new variable the dual periods can be written as follows,
$$
\eqalign{&\Pi_1(x,y) = \Tpi\left(-W(a_1)+(g\Dph^3) {\cal F}(x,y)\right) \cr
        &\Pi_2(x,y) = \Tpi\left( -W(a_2)+(-g\Dph^3) {\cal F}(y,x) \right)}
$$
where,
$$
\eqalign{{\cal F}(x,y) & = x(\log x - 1) -2(x-y)\log({\Lco \over
\Dph})+(2x^2+10xy+5y^2) +
({32\over 3}x^3+91x^2y +\cr 
& 118xy^2+{91\over
3}y^3)+({280\over 3}x^4+{3484\over 3}x^3 y+2636 x^2 y^2+{5272\over
3}xy^3+{871\over 3}y^4)+\ldots}
$$
Note that we have removed the irrelevant constants $W(\Lco)$ in $\Pi_1$
and $W(-\Lco)$ in $\Pi_2$.
Now the effective superpotential is given by,
\eqn\cubeff{
-\Tpi W_{eff}(x,y) = N_1\Pi_1 + N_2\Pi_2 + \alpha g\Dph^3(x-y)}
Let us separate the contributions to \cubeff\ as,
$$
W_{eff}(x,y) = W_{cl}+W_{np}(x,y)
$$
where $W_{cl}=N_1 W(a_1)+N_2 W(a_2)$ and $W_{np}(x,y)=g(\Dph^3)(-N_1
{\cal F}(x,y)+N_2 {\cal F}(y,x))$. In this expression, the cut off
$\Lco$ gets combined with the bare coupling $\alpha$ to generate what
we identify with the gauge theory scale of the underlying $N=2$
$SU(N)$ Yang-Mills theory $\Lambda$ as in \combine.

Having identified the gauge theory scale $\Lambda$ we can proceed to
integrate out $S_1$ and $S_2$ or equivalently $x$ and $y$. The equations
that need to be solved are,
$$
{\partial W_{eff} \over \partial x} =0  \qquad {\partial W_{eff}\over
\partial y}= 0
$$
The leading order can be easily extracted and reads,
$$
N_1 \log(x) = 2 (N_1+N_2)\log\left( {\Lambda \over \Dph}\right) \qquad  N_2
\log(y) = 2 (N_1+N_2)\log\left( {\Lambda \over \Dph}\right)
$$
Now we can see the appearance of the $N_1 N_2$ vacua of the gauge theory
{}from the solutions to the above equations, namely,
$$
x^{N_1} = \left({\Lambda \over \Dph }\right)^{2(N_1+N_2)}  \qquad  y^{N_2} 
= \left({\Lambda \over \Dph }\right)^{2(N_1+N_2)} 
$$
It is useful to define the expansion parameter 
$$
T \equiv \left({\Lambda \over \Dph }\right)^{{2(N_1+N_2)\over N_1 N_2}}
$$
and the solution is then given by 
$$
x = T^{N_2},    \quad  y = T^{N_1}
$$
where the choice of the $N_1 N_2$-th root will determine the vacuum.

Note that the meaning of leading order depends on the values of $N_1$
and $N_2$.  Assuming a power series expansion for $x$ and $y$ in $T$
we can compute order by order $W_{low}$. This gives us the answer
for the $U(N)$ theory. To obtain the answer for $SU(N)$, we only have
to impose that the quantum trace of the chiral superfield be zero:
$\ev{Tr \Phi }={\partial W_{low}(\lambda)\over \partial \lambda} =
0$. This should be imposed order by order in $T$.

\subsec{Quantum tracelessness}

Let us start by writing,
$$
W_{low}(\lambda ,\Lambda )= N_1 W(a_1)+N_2 W(a_2)+g\Dph^3 P(T)
$$
with $P(T) =-N_1 {\cal F}(x_{(T)},y_{(T)})+N_2 {\cal
F}(y_{(T)},x_{(T)})$. It is then easy to see that 
\eqn\quantcon{
{\partial W_{low}(\lambda ,\Lambda)\over \partial \lambda} = N_1 a_1 
+ N_2 a_2 - 2
\Dph \left( 3 P(T)-{2(N_1+N_2)\over N_1 N_2}T {dP(T)\over dT} \right)}
where it was important to remember that $T$ itself depends on
$\lambda$ through $\Dph$. Therefore we are forced to define a better
expansion parameter given by,
$$
t = \left( {\Lambda \over \Dph_c} \right)^{{2(N_1+N_2)\over N_1 N_2}}
$$
where $\Dph_c$ is computed using the Lagrange multiplier obtained by
solving the classical tracelessness constraint,
\eqn\classlam{{\lambda_c\over g} = -{N_1
N_2\over (N_1-N_2)^2}\left({m\over g}\right)^2} Having found $\lambda
=\lambda (t)$ such that the quantum trace \quantcon\ vanishes, we can
use it to compute the low energy superpotential for our $SU(N)$ theory
that is given now as a power expansion in $t$.  It is possible to give
an explicit formula for the first two terms, i.e, the classical
contribution and the first quantum correction for any $N_1$ and
$N_2$. Higher order corrections have to be computed independently in
each case.  Assuming that $N_2 < N_1$, we get,
$$
W_{low}(t) = {1\over 6}{m^3\over g^2}{N_1 N_2 (N_1+N_2)\over
(N_1-N_2)^2}\left[ 1 + {6 (N_1+N_2)^2 \over N_2(N_1-N_2)} t^{N_2} + {\cal
O}(t^{N_2 + 1})   \right].
$$

\subsec{Examples}

Let us consider the different cases for which the deformed $\N =2$
field theory results have been computed in section 6, in order to
compare the answer with that of the geometry.

\vglue 1cm

$U(3N) \rightarrow U(2N)\times U(N)$

\vglue 1cm

We only need to consider the case $U(3)\rightarrow U(2)\times
U(1)$.   As we saw in section 6.1 
this is particularly simple from the field theory perspective,
where $W_{eff}=W_{cl}\pm 2 g\Lambda ^3$, with only one quantum
correction term.  In order to reproduce this simple result, some
miraculous cancellations have to occur order by order in our series.
Since we have computed the dual periods up to order $S_1^4$ and therefore the
effective superpotential up to order $x^4 \sim t^4$, we can not compare
orders equal or higher that $t^5$ even though they already appear in our
computation in the form $x y^2$ or $x^3 y$ since $y\sim t^2$.

Let $N_1 =2 $ and $N_2 = 1$. Integrating out $x$ and $y$ we get,
$$
x_{(T)}= T \left( 1+T+10T^2+140T^3 +\ldots \right) \qquad  y_{(T)}=T^2 
(1+10T+140T^2+\ldots)
$$
Plugging this back in $W_{eff}$ we get the answer for the $U(3)$ case,
$$
W_{low}(T)=W_{cl}+g \Dph^3 (2T+{\cal O}(T^5))
$$
which is consistent with the exact answer $W=W_{cl}+2g \Delta^3 T$
discussed in section 6.1, to the order we have computed.  One might
worry that imposing quantum tracelessness for $SU(3)\rightarrow
SU(2)\times U(1)$ could result in $T$ being a complicated expansion in
terms of $t$. However, one can check that the classical trace is not
corrected quantum mechanically in this case and therefore $T=t$.
We thus have 
$$
W_{low}(t) = {m^3\over g^2}(1+ 54 t+{\cal O}(t^5)),
$$  
and, recalling the definition of $t=\pm\left({g \Lambda \over 3
m}\right)^3$, we get
$$
W_{low}(t)= {m^3\over g^2}\pm 2g\Lambda^3,
$$
in perfect agreement with the field theory result.

We can also use the geometry analysis to obtain the gauge coupling of
the IR $U(1)$ gauge theory photon, and compare with the field theory
analysis.  The field theory result obtained in section 6.1 is that the
original $SU(3)$ curve degenerates as $P_3^2-4\Lambda
^6=(x-m/g)^2F_4(x)$, with $g^2F_4(x)=W'(x)^2\mp 4g^2\Lambda
^3(x+2m/g)$.  The remaining massless photon has gauge coupling given
by the complex modulus $\tau$ of the torus $y^2=F_4(x)$.  This matches
perfectly with the geometry result if, at the extremum of our
effective superpotential for $S$, we have $f_1(x;\ev{S})=\mp
4g^2\Lambda ^3(x+2m/g)$. Strikingly, this is indeed the case.

\vglue 1cm

$U(5N) \rightarrow U(3N)\times U(2N)$

\vglue 1cm
 
In this case the deformed $\N =2$ field theory analysis predicts an
infinite series discussed in section 6.1. From the dual geometry, to
the order we have computed, we will be able to compare up to order
$t^9$ because $x\sim t^2$ and $y\sim t^3$, therefore the $t^{10}$
receives corrections {}from the $x^5$.

Let $N_1=3$ and $N_2=2$. Integrating out $x$ and $y$ we get,
$$
x(T) = T^2 \left(1+{8\over 3}T^2-{10\over 3}T^3 + a_4 T^4+a_5 T^5+a_6 T^6+a_7
T^7+\ldots \right)
$$
and
$$
y(T)= T^3(1+5T^2+11T^3 +b_4T^4+b_5T^5+b_6T^6+\ldots )
$$
The undetermined coefficients are shown to stress the fact that they do not
contribute to the order we are computing, despite being allowed
by naive power counting.
Plugging this back in $W_{eff}$ we get the answer for $U(5)$,
$$
\eqalign{W_{low}(T) = W_{cl}+& g \Dph^3\left(  3T^2-2T^3+4T^4-10T^5+{85\over
3}T^6-{266\over 3}T^7+{8170\over 27}T^8+  \right. \cr
&\left. -{3332\over 3}T^9 +\ldots\right)}.
$$

In this case, we do have to take care with the quantum corrections to
the trace, in order to get the correct $SU(5)$ superpotential.  It
turns out that
$$
{\lambda \over g}= {\lambda_c \over g} \left( 1-{25\over 3}t^2+{100\over
3}t^3-{550\over 3}t^4+{10400\over 9}t^5-7875 t^6+{508300\over
9}t^7-{11338250\over 27}t^8 +\ldots \right).
$$
Using this to compute $T=T(t)$, $a_1=a_1(t)$, and $a_2=a_2(t)$, and plugging
back in the effective superpotential, we get
$$
\eqalign{W_{low}(t) = {250\over 2}{m^3\over g^2}\left( \right.&\left. {1\over
25}+3t^2-2t^3+6t^4-26t^5+135t^6-782t^7+{14630\over 3}t^8 + \right.\cr
& \left. -32076 t^9 +\ldots \right)}.
$$
This is in perfect agreement with the deformed $\N =2$ field theory answer.

\vglue 1cm

$U(5N) \rightarrow U(4N)\times U(N)$

\vglue 1cm

The deformed $\N =2$ field theory analysis again predicts an
infinite series for $W_{eff}$.  Again, this is also seen from 
the geometry dual, and we will be able to compare up to order $t^4$ 
since we have computed the dual periods to order $S_1^4$.
Let $N_1 =4$ and $N_2=1$. Integrating out $x$ and $y$ we get,
$$
x(T)=T\left( 1-{3\over 2}T-{47\over 8}T^2-{73\over 2}T^3+\ldots \right), 
\qquad y(T)=T^4 + {\cal O}(T^5).
$$
Plugging this back in the effective superpotential we get the $U(4)$ answer,
$$
W_{low}= W_{cl}+g\Dph^3 (4T-3T^2-{47\over 6}T^3-{75\over 2}T^4+\ldots).
$$
For the $SU(4)$ case, the vanishing of the quantum corrected trace
implies that,
$$
{\lambda \over g}= {\lambda_c \over g}\left(1+{25\over 3}t+{25\over
9}t^2+{175\over 72}t^3 +\ldots\right).
$$
Using this as in the previous case, we finally get the low energy
superpotential to be
$$
W_{low}={125\over 27}{m^3\over g^2}\left( {2\over 25}+4t-{1\over 3}t^2-{7\over
54}t^3-{5\over 54}t^4 + {\cal O}(t^5)\right).
$$
This exactly agrees, to this order, with the expected answer.

\bigskip
\centerline{{\bf Acknowledgments}}

We would like to thank S. Katz for participation at the initial
stages of this project.  We would also like to thank
M. Aganagic, J. Edelstein and K. Hori for valuable discussions.
 
 The work of F.C. and C.V. is supported in part by NSF grants PHY-9802709
and DMS 9709694.
The work of K.I. is supported by DOE-FG03-97ER40546.

\appendix{A}{Deformed $\N =2$ field theory analysis 
for $U(5)\rightarrow U(3)\times U(2)$}

We here find the supersymmetric vacua of the deformed $\N =2$ theory 
for one of
the splittings of $U(5)$. The analysis goes along much the same lines as for
$SU(5)$. We parameterize
\eqn\unpoly{
P_5(x) = (x-(q+a))^2(x-(q+b))^2(x-(q+c))\mp 1.}  For $SU(5)$, $q$ was
fixed by the tracelessness condition but now it is a free parameter.
Since $a$ and $b$ appear in a symmetric way it turns out to be useful
to define $s=a+b+2q$ and $k=(a+q)(b+q)$. The constraints are now given
by,
$$
k = q^2-q(2 q-s)+2 c(2 q-s)  \qquad  4(2q-s)^2c^3=\pm 1
$$
{}From \unpoly\ we can read off $u_1$, $u_2$, and $u_3$ using that,
$$
P_5(x) = x^5 - u_1 x^4+({1\over 2}u_1^2-u_2)x^3+({1\over 6}u_1^3+u_1
u_2-u_3)x^2+\ldots
$$
Plugging $u_i=u_i(q,s,c)$ in $W_{eff}$ and introducing a Lagrange
multiplier in order to impose the constraint left after we eliminate $k$,
we get,
$$
W_{eff}=g u_3+mu_2+\lambda u_1+h(\mp \Lambda^5-4(2q-s)^2c^3)
$$
The equations we need to solve are given by $\partial W_{eff}/\partial
c=0$, $\partial W_{eff}/\partial s=0$, and $\partial W_{eff}/\partial
q=0$. Using the first to eliminate $h$ in the second and the third,
these equations simplify to,
$$
\lambda + m (-q+c+s)+g(q^2+c^2+3cs+s^2-2q(2c+s))=0
$$
$$
-5\lambda - m(q+c+2s) - g(5q^2-14qc+c^2-4qs+8cs+2s^2)=0
$$
$$
4(2q-s)^2c^3=\pm \Lambda^5
$$

In order to find an expansion parameter around the classical solution we have
to take the limit $\Lambda=0$ and solve the equations. We find that,
$$
q={-m+\sqrt{m^2-4g\lambda}\over 2g} \quad  s=-{m\over g}
$$
Therefore, $(2q-s) = \sqrt{{m^2\over g^2}-4{\lambda\over g}}=\Delta$ and it is
clear that the expansion parameter is $T$ given by $T^6=({\Lambda \over
\Delta })^{10}$. Again, there are six possible solutions giving the six possible
vacua $N_1 N_2=6$, since $N_1=2$ and $N_2=3$.

Solving these equations assuming a power expansion in $T$ for $s=s(T)$,
$q=q(T)$ and $c=c(T)$, we get after plugging back in the effective
superpotential,
$$
W_{low} = 3 W(a_1)+2 W(a_2)+ g \Delta \left(3T^2+2T^3+4T^4+10T^5+\ldots \right)
$$
where $W(x)={g\over 3}x^3+{m\over 2}x^2+\lambda x$, $W'(x)=g(x-a_1)(x-a_2)$
and $\Delta = a_1-a_2$.

\appendix{B}{Computation of Periods for the cubic superpotential}

In this appendix we will show the explicit computation of the corrections
$P(S_1,S_2)$ in the expression for $\Pi_1$ in \dualone. 

The computation of $P(S_1,S_2)$ will not be done directly in
terms of $S_1$ and $S_2$, we will write all four periods in terms
of two new variables $\Dd$ and $\Dc$ - to be defined below - and at the end
we will recollect $P(S_1,S_2)$. This procedure can be done
systematically up to any order in $S_i$'s.

{\bf Computation}:

For practical purposes we will write the effective one-form as follows
\eqn\oneform{dx\sqrt{W'^2(x)+f_1(x)}=dx\
g\sqrt{(x-x_1)(x-x_2)(x-x_3)(x-x_4)}}

It is also convenient to define new variables given by, 
$$
\Dd \equiv {1\over 2}(x_2 - x_1)  \quad \Dc \equiv {1\over 2}(x_4-x_3)
$$
$$
Q \equiv {1\over 2}(x_1+x_2+x_3+x_4)  \quad  I \equiv {1\over 2}((x_3+x_4)-(x_1+x_2))
$$

It is clear that since $f_1(x)$ is considered a small
perturbation we will have  
$$\mid \Dd\mid \sim \mid
\Dc\mid \ll \mid I \mid .$$
 We will use this in order to expand all 
four periods in
powers of $\Dd$ and $\Dc$.

Let us consider $S_1$. For this we change variables to $y=x-{1\over
2}(x_1+x_2)$ and the integral becomes:
$$
S_1 ={g\over 2\pi }\int^{y_4}_{y_3}\sqrt{(y-y_3)(y-y_4)}\sqrt{y^2-\Dd^2}
$$
Expanding the second square root for $\Dd$ small, each term in the series
can be computed explicitly and it is most easily given in terms of a generating
function,
\eqn\generone{F(a) \equiv -\pi \sqrt{(y_3+a)(y_4+a)}+{\pi\over 2}(y_3+y_4+2a)}
as follows,
$$
S_1 = {g \over 32}(y_3+y_4)(y_4-y_3)^2 +{g\over 2\pi}\sum^{\infty}_{n=1}c_n
\Dd^{2n} F^{(n)}(0) 
$$
where $c_n$ are the coefficients in the expansion of $\sqrt{1-x}$ and
$F^{(n)}(a)$ is the $n$-th derivative with respect to $a$.

The explicit answer has the following structure,
\eqn\funone{
S_1 = {g\over 4}\Dc^2 I - {g\over {2I}}K(\Dd^2,\Dc^2,I^2)}
where
$$
K(x,y,z) = {1\over 4}xy\left(1+{1\over 4z}(x+y)+{1\over
8z^2}(x+y)^2+{1\over 8z^2}xy+\ldots \right)
$$

It is important to notice that this is symmetric in $(x,y)$, namely,
$K(x,y,z)=K(y,x,z)$. This allows us to write,
\eqn\funtwo{
S_2 = -{g\over 4}\Dd^2 I + {g\over {2I}}K(\Dd^2,\Dc^2,I^2)}

Let us now compute the dual periods starting with $\Pi_1$. In this
case we can use the same expansion as before for $S_1$, however, we
have to keep in mind that $\Lco$ will be taken to infinity at the end
and therefore we shall discard any contribution of order $\Lco^{-1}$
or higher in an expansion around infinity.

In this case it is also useful to define a generating function,
\eqn\genetwo{
G(a) = \sqrt{(I+a)^2-\Dc^2}
\log\left({\sqrt{(I+a)+\Dc}+\sqrt{(I+a)-\Dc}\over
\sqrt{(I+a)+\Dc}-\sqrt{(I+a)-\Dc}}\right)} and the answer is given by,
\eqn\finpone{
\eqalign{{2\pi i\over g}\Pi_1 =& {1\over 3}\Lco^3-{1\over 2}Q \Lco^2 + {1\over
4}(Q^2-I^2-2(\Dc^2+\Dd^2))\Lco + {1\over 2}I(\Dd^2-\Dc^2)\log \Lco \cr
& -{1\over 24}(I+Q)^3+{1\over
8}I(I+Q)^2+{1\over 4}\Dd^2 (I+Q)+{1\over 4}\Dc^2 Q+ \cr 
& {1\over 2}I(\Dc^2-\Dd^2)\log(2\Dc) +\sum^{\infty}_{n=1}c_n
\Dd^{2n} G^{(n)}(0)}}
where $c_n$ are as before the coefficients of the power expansion of
$\sqrt{1-x}$.

This result is not enough because we want it to show only explicit
dependence on the classical superpotential parameters $m,g,\lambda$ and the
two deformation parameters $\Dd$ and $\Dc$. In order to do this we only
have to realize that since $f_1(x)$ in \oneform\ is of degree one and
$W'^2(x)$ of degree four, then the coefficients of $x^3$ and $x^2$ are given
in terms of the classical roots $a_1$ and $a_2$. This allows us to write,  
$$
Q =a_1+a_2  \qquad  I^2 = (a_1-a_2)^2-2(\Dd^2+\Dc^2)
$$

Using this, \funone\ and \funtwo\ we can explicitly compare order by
order in $\Dd$ and $\Dc$ the two expressions for $\Pi_1$ given by
\finpone\ and \dualone\ to obtain the following result,
$$
\eqalign{2\pi i\;\Pi_1 =&W(\Lco)-W(a_1)+S_1 (\log{S_1\over g\Dph} - 1) 
+ 2S_2\log\Dph - 2
(S_1+S_2)\log \Lco + \cr &+ g(\Dph)^3 \left[{1\over (g\Dph^3)^2}\left(
2S_1^2-10S_1 S_2+5 S_2^2\right) + {1\over (g\Dph^3)^3}\left( {32\over
3}S_1^3-91S_1^2S_2 +\right. \right.\cr &\left. +118S_1S_2^2-{91\over
3}S_2^3\right) + {1\over (g\Dph^3)^4}\left({280\over
3}S_1^4-{3484\over 3}S_1^3S_2+2636S_1^2S_2^2+\right. \cr
&\left.\left. -{5272\over 3}S_1S_2^3+{871\over 3}S_2^4 \right)+ {\cal
O}\left({S^5 \over (g\Dph^3)^5}\right) \right].}
$$

Likewise we can get $\Pi_2$ from the above result by simply exchanging
$a_1 \leftrightarrow a_2$, $S_1 \leftrightarrow S_2$, $\Dph \leftrightarrow
-\Dph$ and $\Lco \leftrightarrow -\Lco$. This leads to,
$$
\eqalign{2\pi i\;\Pi_2 =&W(-\Lco)-W(a_2)+S_2 (\log{S_2\over g\Dph } - 1) 
+ 2 S_1\log\Dph - 2
(S_1+S_2)\log \Lco + \cr
&- g(\Dph)^3 \left[{1\over (g\Dph^3)^2}\left( 2S_2^2-10S_1 S_2+5 S_1^2\right)
- {1\over (g\Dph^3)^3}\left( {32\over 3}S_2^3-91S_2^2S_1 +\right. \right.\cr
&\left. +118S_2S_1^2-{91\over
3}S_1^3\right) 
 + {1\over (g\Dph^3)^4}\left({280\over 3}S_2^4-{3484\over
3}S_2^3S_1+2636S_2^2S_1^2+\right. \cr
&\left.\left. -{5272\over 3}S_2S_1^3+{871\over 3}S_1^4 \right)+
{\cal O}\left({S^5 \over (g\Dph^3)^5}\right) \right]} 
$$ 

This completes our computation of the periods.

\listrefs
\end